\documentclass[lettersize,journal]{IEEEtran}
\usepackage{amsmath,amsfonts}
\usepackage{algorithmic}
\usepackage{algorithm}

\usepackage{graphicx}
\usepackage{subfigure} 
\usepackage{gensymb}
\usepackage{booktabs}
\usepackage{url}
\usepackage{color}
\usepackage{amsmath}
\definecolor{brickblue}{rgb}{0.2,0.25,0.93}

\begin{document}

\title{$Radar^2$: Passive Spy Radar Detection and Localization using COTS mmWave Radar}

\author{Yanlong Qiu, Jiaxi Zhang, Yanjiao Chen, Jin Zhang,~and~Bo~Ji
}



\maketitle

\begin{abstract}
Millimeter-wave (mmWave) radars have found applications in a wide range of domains, including human tracking, health monitoring, and autonomous driving, for their unobtrusive nature and high range accuracy. These capabilities, however, if used for malicious purposes, could also result in serious security and privacy issues. For example, a user's daily life could be secretly monitored by a spy radar. Hence, there is a strong urge to develop systems that can detect and locate such spy radars. In this paper, we propose $Radar^2$, a practical system for passive spy radar detection and localization using a single commercial off-the-shelf (COTS) mmWave radar. Specifically, we propose a novel \textit{Frequency Component Detection} method to detect the existence of mmWave signals, distinguish between mmWave radar and WiGig signals using a waveform classifier based on a convolutional neural network (CNN), and localize spy radars using triangulation based on the detector's observations at multiple anchor points. Not only does $Radar^2$ work for different types of mmWave radar, but it can also detect and localize multiple radars simultaneously. Finally, we performed extensive experiments to evaluate the effectiveness and robustness of $Radar^2$ in various settings. Our evaluation results show that the radar detection rate is above 96$\%$ and the localization error is within 0.3m. The results also reveal that $Radar^2$ is robust against various environmental factors (e.g., room layout and human activities). 
\end{abstract}

\vspace{30pt}

\section{Introduction}

\IEEEPARstart{M}illimeter-wave (mmWave) radars have gained popularity in recent years due to the high resolution provided by the wide bandwidth. They have been explored in a wide variety of sensing systems, such as human tracking \cite{zeng2016human,huang2021indoor}, health monitoring \cite{wang2020remote,yang2017vital,alizadeh2019remote, guo2021design}, and autonomous vehicles \cite{mmWaveRadarAV}. Compared with wearable devices or cameras, mmWave radars are less intrusive and more robust. Besides, it achieves a higher range of resolution for accurate localization.

In particular, mmWave radar systems have been widely used for detecting users' location, activity, and vital signs in an unobtrusive manner \cite{yue2020bodycompass,fan2020home,wang2020remote,yang2017vital}. However, this ``unobtrusiveness'' feature can also pose a threat to users. For example, Zeng et al. \cite{zeng2016human} indicated that unsolicited radars can be used to spy on users' daily lives without users' awareness. Moreover, mmWave radars can even spy on screens or computers, which may cause information leakage \cite{LED, Li2022SpiralSpy}. With more radar systems being developed to monitor sensitive information, the wide adoption of mmWave radars can introduce serious security and privacy concerns at the same time. Therefore, it is of critical importance to detect malicious spy radars. It is also imperative to locate the radar(s) and act accordingly once they are successfully detected. To the best of our knowledge, current mmWave radar actively transmits signals.

\begin{figure*}[htbp]
    \centering
    \includegraphics[width=250pt]{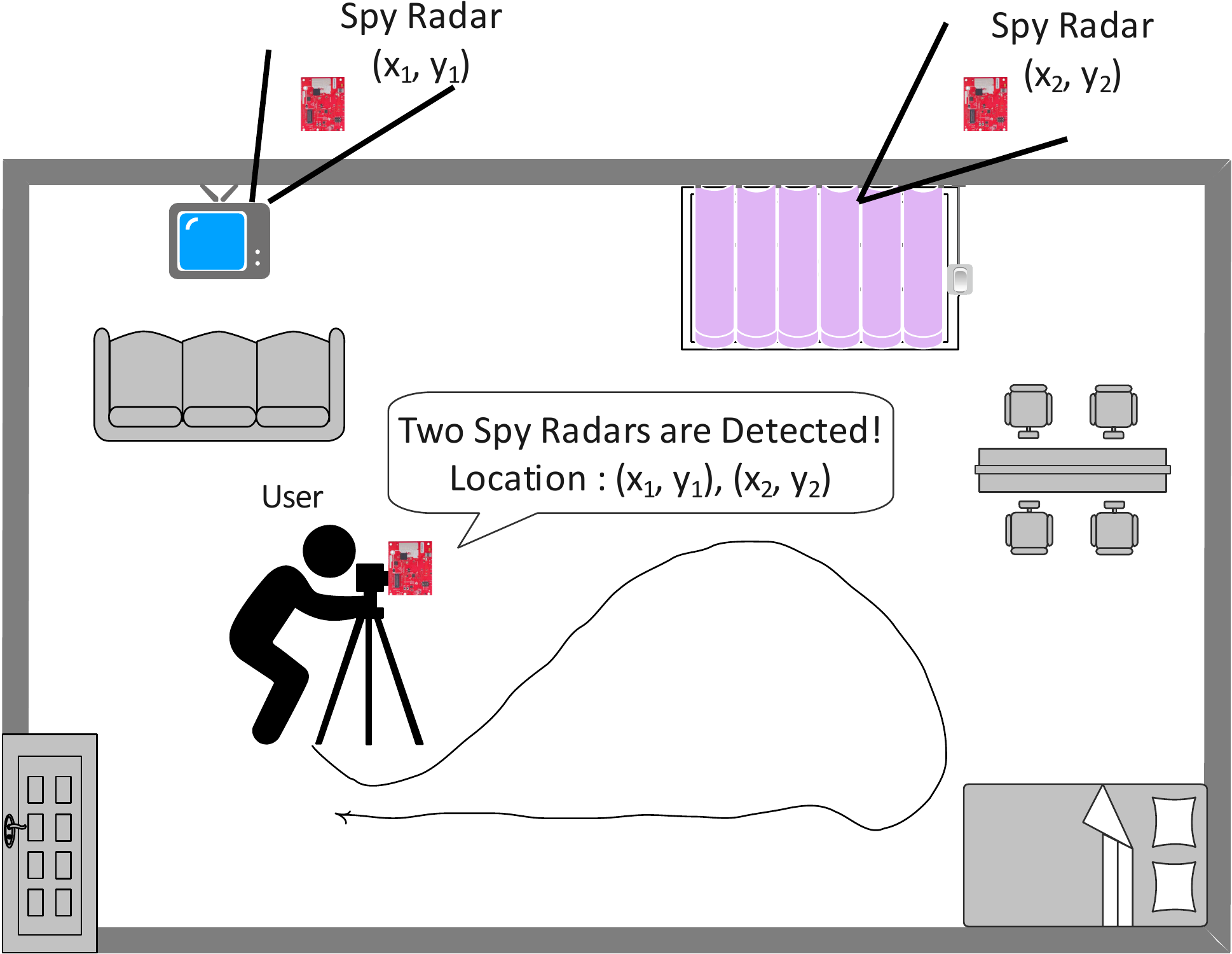}
    \caption{An application scenario of $Radar^2$. The user leverages a mmWave radar to detect the presence of other mmWave radars and localize them by measuring the signals at multiple anchor points.}
    \label{fig:scene}
\end{figure*}

It has been proposed that spectrum analyzers can be used to detect mmWave signals \cite{acharya2013detection, weng2005neural, acharya2012system}. In 2012, Shikhar et al. proposed that unintended electromagnetic emissions (UEEs) can be taken as a unique signature of electronic devices, and such a signature can be used for device detection and identification \cite{acharya2012system}. However, a spectrum analyzer is required to sample UEEs. A key limitation of such systems is the high cost. The spectrum analyzer needs to have a high sampling rate to detect mmWave radar signals. For example, to recover mmWave signals with a bandwidth of 4GHz in the 77-81GHz frequency band, the required sampling rate is usually 8GHz according to the Nyquist Theorem \cite{gruber2013proofs}. Spectrum analyzers with such a high sampling rate are often costly and unsuitable for daily use. 

Furthermore, even though the spectrum analyzer can detect the existence of mmWave signals \cite{acharya2013detection, weng2005neural, acharya2012system}, it cannot localize the hidden radar. The localization of the radar is crucial to enabling users to locate and remove it. To overcome the limitations of spectrum analyzers, we aim to use an inexpensive, commercial off-the-shelf (COTS) mmWave radar to detect the presence of spy radars and locate them.

We propose $Radar^2$, a practical spy radar detection and localization system using a COTS mmWave radar. In comparison with previous work using spectrum analyzers \cite{acharya2013detection, weng2005neural, acharya2012system}, $Radar^2$ can not only detect the spy radar more affordably but also be able to localize the spy radar. The application scenario is shown in Fig. \ref{fig:scene}, where spy radars are installed in imperceptible places (e.g., behind the curtain or TV). $Radar^2$ can detect spy radars using a hand-held COTS radar as the detector and alarm the user. To locate spy radars, the user should hold the detector and walk to several positions to measure the signals. Then, the locations of spy radars can be calculated and reported to the user. 

To design such a system, we face the following challenges: \textit{First, the hardware constraint of COTS mmWave radars introduces new challenges for radar detection}: 1) COTS mmWave radars have a limited sampling rate, which is not enough to directly recover other mmWave signals (which a spectrum analyzer does). 2) Since COTS mmWave radars are intended for sense, they have a predefined processing flow and limited flexibility. The predefined signal processing procedures are designed to derive the sensing object's distance and velocity instead of detecting and recovering signals from another radar. Therefore, it is unclear how to detect signals from spy radars using a COTS mmWave radar with such hardware constraints.

\textit{Second, other wireless systems, such as WiGig for 60GHz wireless network \cite{hansen2011wigig}, also function in the mmWave band.} Therefore, we need to be able to distinguish radar signals from such WiGig signals working in the same band. 

\textit{Thirdly, there is no mature solution for localizing spy radars.} Even though mmWave radars can localize objects, they cannot distinguish a spy radar from other objects. Thus, conventional localization methods for mmWave radar, like FFT, cannot be used to locate spy radars. In addition, if multiple spy radars work simultaneously, detecting their presence and locating them becomes more challenging since their signals could be superimposed on each other.

To overcome the aforementioned challenges, we propose $Radar^2$ for mmWave signal detection, classification, and radar localization. To detect radars' presence, we design a \textit{Frequency Component Detection} method to detect mmWave signals under the constraint of the current COTS radar processing flow. Specifically, we demodulate the received signal using a sweep frequency signal and multiple single-frequency signals. Only when the received signal and the carrier signal have the same frequency components simultaneously can the intermediate frequency (IF) signal be sampled. Otherwise, the frequency component in the IF signal will be filtered by a low-pass filter, for non-destructive sampling \cite{gruber2013proofs}. Therefore, we can observe a peak in the baseband if the mmWave signal exists. This peak refers to the time when the received signal and the carrier are at the same frequency. By finding peaks in the demodulated signal, we can detect the existence of mmWave signals.

However, another challenge is the existence of WiGig systems that function in the same mmWave band. Therefore, we must design a waveform classifier to distinguish radar signals from WiGig signals. By observing that WiGig and radar signals have different spectrums, we design a CNN-based classifier to distinguish WiGig signals from radar signals.

To localize a spy radar using only one detector, we propose a solution based on \textit{Triangulation} by measuring signals at multiple positions. Note that a key difference between spy radars and other objects is that spy radars actively transmit signals. Therefore, we can estimate the Angle-of-Arrival (AoA) of the received signals at several anchor points and combine these observations to estimate the target location. We first estimate the AoA using Multiple Signal Classification (MUSIC) at each anchor point and then move our detector to different positions to repeat the measurements. Finally, we design an \textit{nearest approaches} to localize the spy radar by exploiting multiple AoAs observed at different positions.

When multiple devices exist, the detector receives a combination of these signals. We assume that these devices are located at different positions. To distinguish and identify these signals, we propose to separate them by their unique AoAs. We extract them one by one for the detection and localization of multiple spy radars.

We summarize our contributions as follows:

\begin{itemize}

    \item We propose $Radar^2$, a practical passive spy radar detection and localization system using a single COTS mmWave radar. To the best of our knowledge, this is the first work on the detection and localization of spy radars. 

    \item To realize $Radar^2$, we propose a novel \textit{Frequency Component Detection} method to detect the existence of mmWave signals, distinguish between mmWave radar and WiGig signals using a waveform classifier based on a convolutional neural network (CNN), and localize spy radars using triangulation based on the detector's observations at multiple anchor points. Not only does $Radar^2$ work for different types of mmWave radar, but it can also detect and localize multiple radars simultaneously.

    \item We performed extensive experiments to evaluate the effectiveness and robustness of $Radar^2$ in various settings. Our evaluation shows that the radar detection rate is above 96$\%$, and the localization error is within 0.3m. The results also reveal that $Radar^2$ is robust against various environmental factors (e.g., room layout and human activities).

\end{itemize}

\section{Threat Model}

We consider the scenario where a victim, referred to as Alice, is monitored by an attacker, Bob, using mmWave radar remotely in the same room or outside the wall. Bob installs a spy radar in Alice's room or outside the wall, and uses the spy radar to generate mmWave and receive the reflected signal, to spy on Alice and obtain various information:

\textit{Activity and location information:} A hidden spy radar can provide the location and activity of Alice to Bob. Bob could plan a home invasion based on Alice's activity and location.

\textit{Vital sign information:} mmWave radar can sense vital signs like breathing or heartbeat \cite{wang2020remote,yang2017vital,alizadeh2019remote, guo2021design}, thus Bob can get access to the healthcare data about Alice. Bob could even sell the healthcare data to other companies such as Apple and Google. This would raise the price of Alice's health plan, which could lead to Alice losing her property.

\textit{Screen and account information:} Moreover, unlike spy cameras, mmWave radar can work through the wall and spy on screens and computers \cite{LED, Li2022SpiralSpy}. When Alice is surfing the Internet, a spy radar may be able to deduce private information like login password, voice token, and credential image.

These attacks will raise privacy concerns, and cause data leakage for Alice. To prevent such spy radar attacks, a detection system should be developed for Alice. This detection system should satisfy the following requirements.

\textbf{Availability.} The cost of the system should be acceptable, and the detection device should be portable.

\textbf{Localization.} Except for the detection of spy radars, the system should also be able to locate the spy radars.

\textbf{No Transmission.} Our detection system should not emit any radio signal, otherwise, the attacker may observe the existence of the detector and pose more serious threats. Therefore, the detector should be able to detect the existence of spy radars passively, i.e., disable its transmission.

\section{Background}
\label{background}

Millimeter wave radars transmit electromagnetic signals in the mmWave band with frequencies ranging from 30 to 300 GHz and detect the movement of objects based on the reflected signal. According to different patterns of transmitted signals, mmWave radars can be classified into different categories. Therefore, our system should be able to detect the presence of all various types of radar. Besides, we aim at using COTS radar to implement $Radar^2$. Therefore, our design should be under the constraint of the current radar processing flow. Meanwhile, we need to differentiate mmWave radar with a 60GHz wireless transceiver WiGig \cite{60GHz}, which also works in the mmWave band.

\subsection{Type of mmWave Radar}
There are various kinds of mmWave radars for different applications. We will discuss the features and applications of commonly used radars in this section.

\subsubsection{CW Radar}
Continuous wave (CW) radar transmits a continuous electromagnetic wave signal with constant amplitude and frequency, typically a sine wave $s(t)$:
\begin{equation}
    s(t)=e^{j2\pi f_0 t+\phi},
\end{equation}
where $f_0$ is the carrier frequency, and $\phi$ is the initial phase. The receiver demodulates the signal to the baseband and measures the velocity $v$ based on the Doppler frequency shift in $f_0$ as
\begin{equation}
    v=\frac{f_d}{2\times f_0}\times c,
\label{equaV}
\end{equation}
where $f_d$ is the Doppler frequency of the target measured by the radar, and $c$ is the propagation speed of the electromagnetic wave.

CW radar can measure velocity and relative distance change but cannot measure absolute distance. Therefore, CW radar is usually used for motion, speed, vibration, and vital signs detection.

\subsubsection{Pulse Radar}

The pulse radar emits short and powerful pulses during the silent period and then receives echo signals. Pulse radar can estimate absolute range by the time delay calculated from the echo signal. They are designed mainly for long-distance sensing. It is primarily used in the military. Other applications include air traffic control, weather observation, and satellite-based remote sensing of the earth's surface.

\subsubsection{FSK Radar}

Frequency Shift Keying (FSK) radar transmits two or more signals with different carrier frequencies alternatively. The transmitted signal $s(t)$ is:

\begin{equation}
    s(t)=\left \{
              \begin{array}{cc}
                e^{2\pi f_a t +\theta_1}, 0<t\leq\frac{T}{2},   &  \\
                e^{2\pi f_b t +\theta_2}, \frac{T}{2}<t\leq T   & 
              \end{array}
    \right.
\end{equation}
where $f_a$ and $f_b$ are two different carrier frequencies, $\theta_1$ and $\theta_2$ are the initial phases of the two signals, and $T$ is the period of FSK signal. 

Suppose there is an object located at a distance $d_0$, then the absolute distance between the target and the radar can be calculated as follows:
\begin{equation}
    d_0 = \frac{c(\Delta \theta_a-\Delta \theta_b)}{4\pi(f_a-f_b)},
\end{equation}
where $\Delta\theta_a, \Delta\theta_b$ are the phase difference between the transmitted signal and received signals at each frequency.

The range can be determined using the phase difference between two echo signals. However, FSK radar cannot differentiate multiple objects at different distances. Therefore, FSK radar can only be used to detect the range and velocity of a single object.

\subsubsection{FMCW Radar}

FMCW radar transmits linearly frequency-modulated signals over time, and the most commonly used signal is a chirp signal. 

The FMCW radar transmits a chirp $x(t)$:
\begin{equation}
\label{TransmitSig}
    x(t)=e^{j(2\pi f_{L}t+\pi S t^2)},
\end{equation}
where $f_L$ is the start frequency and $S$ is the frequency slope.

Let us suppose that there is a certain object at a distance $d_i$ from the radar. Once the receiver receives the signal, it dechirps the received signal with the transmitted signal to obtain the intermediate frequency (IF) signal $y(t)$:

\begin{equation}
    y(t)=r(t)*\overline{x(t)}=\sum_i\alpha_i e^{j(2\pi f_{L}\tau_i-\pi S\tau_i^2-2\pi S\tau_it)},
\end{equation}
whose frequency depends on the time delay $\tau_i$, $f=-2\pi S\tau_i$. Fast Fourier Transform (FFT) of the IF signal can be used by FMCW radars to determine the range.

FMCW radar can detect both the distance and the speed of targets. Besides, FMCW radar can detect multiple objects at different distances, and it is widely used for remote sensing due to its high resolution in range and speed measurement. Typical applications for FMCW radar include indoor navigation, gesture recognition, and vital sign monitoring \cite{FMCWApplication}.

\subsection{Radar Processing Flow}
\label{radarproflow}

Our system should be able to detect the existence of spy radar using COTS mmWave radar. Unlike mmWave band spectrum analyzers that have a high enough sampling rate to recover Radio Frequency (RF) signals, a COTS mmWave radar is hard to detect another mmWave signal since a low-pass filter is used for nondestructive sampling. Furthermore, our design should be within the limits of the COTS mmWave radar's processing flow. Therefore, we need a deeper understanding of its processing flow.

\begin{figure}[t]
    \centering
    \includegraphics[width=200pt]{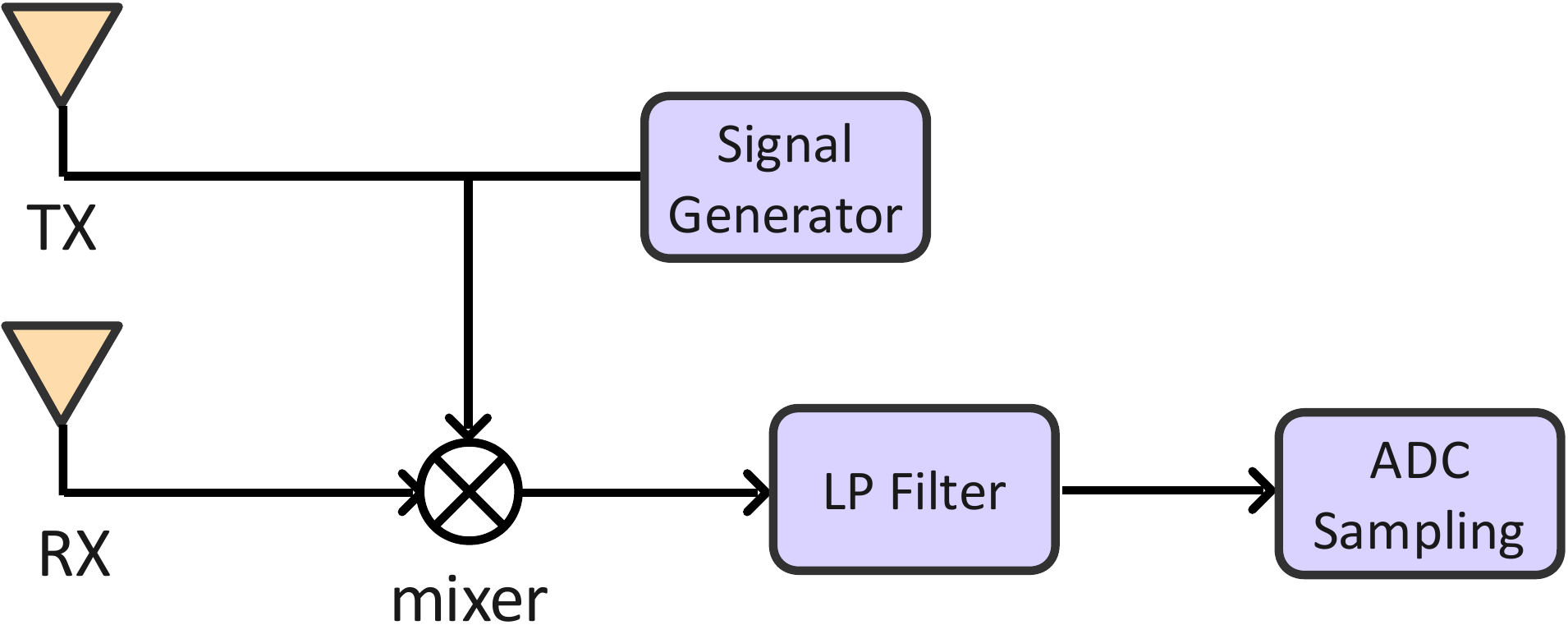}
    \caption{Signal processing flow of COTS mmWave radar.}
    \label{fig:flow}
\end{figure}

The processing flow of a COTS mmWave radar is shown in Fig. \ref{fig:flow}. To sample the received signal using a low sampling rate, the receiver usually mixes the received and transmitted signals in the baseband. The mixer performs conjugation and multiplication of two signals, and we will use demodulation to refer to this operation in the following section. After demodulation, the receiver applies a low-pass filter to remove high-frequency components, which guarantees that the low-sampling rate ADC can sample the IF signal non-destructively. The cutoff frequency of the low pass filter is usually less than half of the sampling rate. Therefore if the frequency difference between the received and transmitted signal is over the cutoff frequency, the receiver is not able to sample the IF signal. This brings challenges to spy radar detection.

\subsection{WiGig}

WiGig is a wireless standard that uses a 60GHz band to provide high-speed wireless transmission. Currently, there are two standards for WiGig, which are IEEE 802.11ad and IEEE 802.11ay. Both use Orthogonal Frequency Division Multiplexing (OFDM) as a physical layer modulation scheme to exploit the benefit of the wide bandwidth of the 60GHz channel. As WiGig can provide as high as 176 Gbps data rate, it will be widely adopted for high-speed video transmission in the home environment. WiGig transceivers are not radars but work in the same frequency band as mmWave radar. Therefore, our system should be able to differentiate WiGig signal from radar signal by analyzing its signal pattern to avoid false detection.

\section{System Design}
\label{System}

The goal of $Radar^2$ is to detect the presence and location of hidden spy radars using a COTS radar. We design $Radar^2$ to achieve the following targets: (1) \textit{radar presence detection}: detect whether there exists a spy radar. (2) \textit{spy radar localization}: accurately localize the position of hidden spy radars using only one detector. (3) \textit{multi-device detection}: detect the presence and location of spy radars when multiple mmWave devices exist simultaneously.

\begin{figure*}[htbp]
    \centering
    \includegraphics[width=400pt]{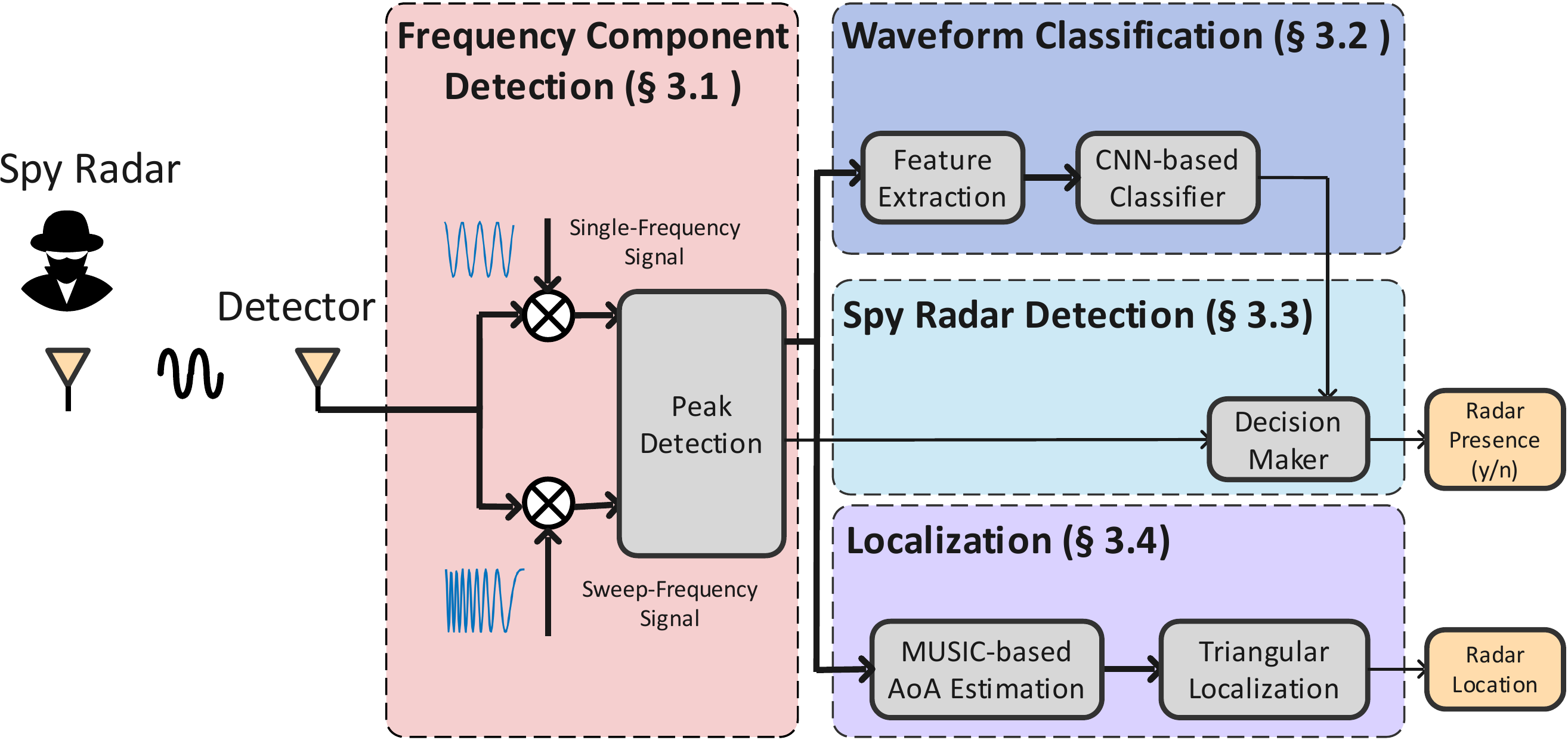}
    \caption{System architecture of $Radar^2$.}
    \label{fig:System Overview}
\end{figure*}

The system architecture of $Radar^2$ is shown in Fig. \ref{fig:System Overview}. Assuming there is a spy radar generating mmWave radar signals in space. The detector of our $Radar^2$ system is also a COTS mmWave radar. The detector's antennas receive the RF signals in space and are passed into the detection system, which is also implemented on the COTS radar board. The detection system comprises four parts: (1) Frequency Component Detection. We propose a module \textit{Frequency Component Detection} to detect the existence of a mmWave signal with a specified frequency in the mmWave band. (2) Waveform Classification. Since there exist other mmWave signals like the WiGig signal, which is not a radar signal. Therefore, we design a waveform classifier to differentiate WiGig from spy radars based on its signal pattern. (3) Spy Radar Detection. We exploit the decision maker, which combines the results of Frequency Component Detection and Waveform Classification, to decide whether there exists a spy radar or not. (4) Localization. We use \textit{Triangulation} to localize the spy radar. Specifically, we move our detector to different positions and calculate the angle of arrival (AoA) of the spy radar using the MUSIC algorithm at each position. With the knowledge of multiple known positions of the detector and its corresponding AoA observations, the position of the spy radar can be calculated.

In our $Radar^2$ system, the Frequency Component Detection Module is the basis of the whole system. In this module, we creatively proposed the signal modulation method and designed two signal patterns that perfectly leverage the mmWave radar signal processing flow. This allows us to detect whether a certain frequency component exists in the mmWave band. We will describe the detailed design of Frequency Component Detection in the following subsections.  

\subsection{Frequency Component Detection}
\label{fcd}

To detect the presence of a spy radar, we have to detect the existence of a signal in the mmWave band first. The key idea of \textit{Frequency Component Detection} is to detect whether there exists a signal component with a specified frequency $f_0$ in the received RF signal ($f_0$ is set to be within the mmWave band to detect mmWave radar), leveraging the processing flow of a mmWave radar. 

COTS mmWave has a given processing flow that we have discussed in Section \ref{radarproflow}. Therefore, detecting the existence of other mmWave radars is challenging. Under the constraint of the given processing flow, we expect to detect the existence of mmWave signal sent by the spy radar using a COTS radar.

To detect whether there exists a signal component with a specified frequency $f_0$ in the received RF signal, using the COTS radar's signal processing flow, we designed the frequency component detection as follows. The detecting radar generates a signal of frequency $f_0$ and multiplies this signal with the received RF signal (spy radar signal) and passes the multiplication result with a low-pass filter, and then checks whether there is a DC component. If there is a DC component, then the received RF signal includes a frequency component of $f_0$. The processing flow is the same as a COTS radar processing flow shown in Fig. \ref{fig:flow}. 

Assuming the received signal is a single frequency wave with $f_0$, 
\begin{equation}
    r(t)=e^{j(2\pi f_0 t)},
\end{equation}
where $f_0$ is the unknown single frequency, and the objective is to demodulate the signal to the baseband so that the ADC can sample on it. It is impossible to traverse all frequencies in the mmWave band to find such a frequency $f_0$, so we propose generating a sweep-frequency signal that sweeps from the lowest frequency to the highest frequency in the mmWave band and using this signal to multiply with the received signal.

The frequency in the sweep-frequency signal can be written as $f_t=at+b$, where the frequency changes with time. The sweep-frequency signal can be written as
\begin{equation}
    c(t)=e^{j(2\pi f_t t)}.
\end{equation}
\begin{figure}[t]
    \centering
    \includegraphics[width=240pt]{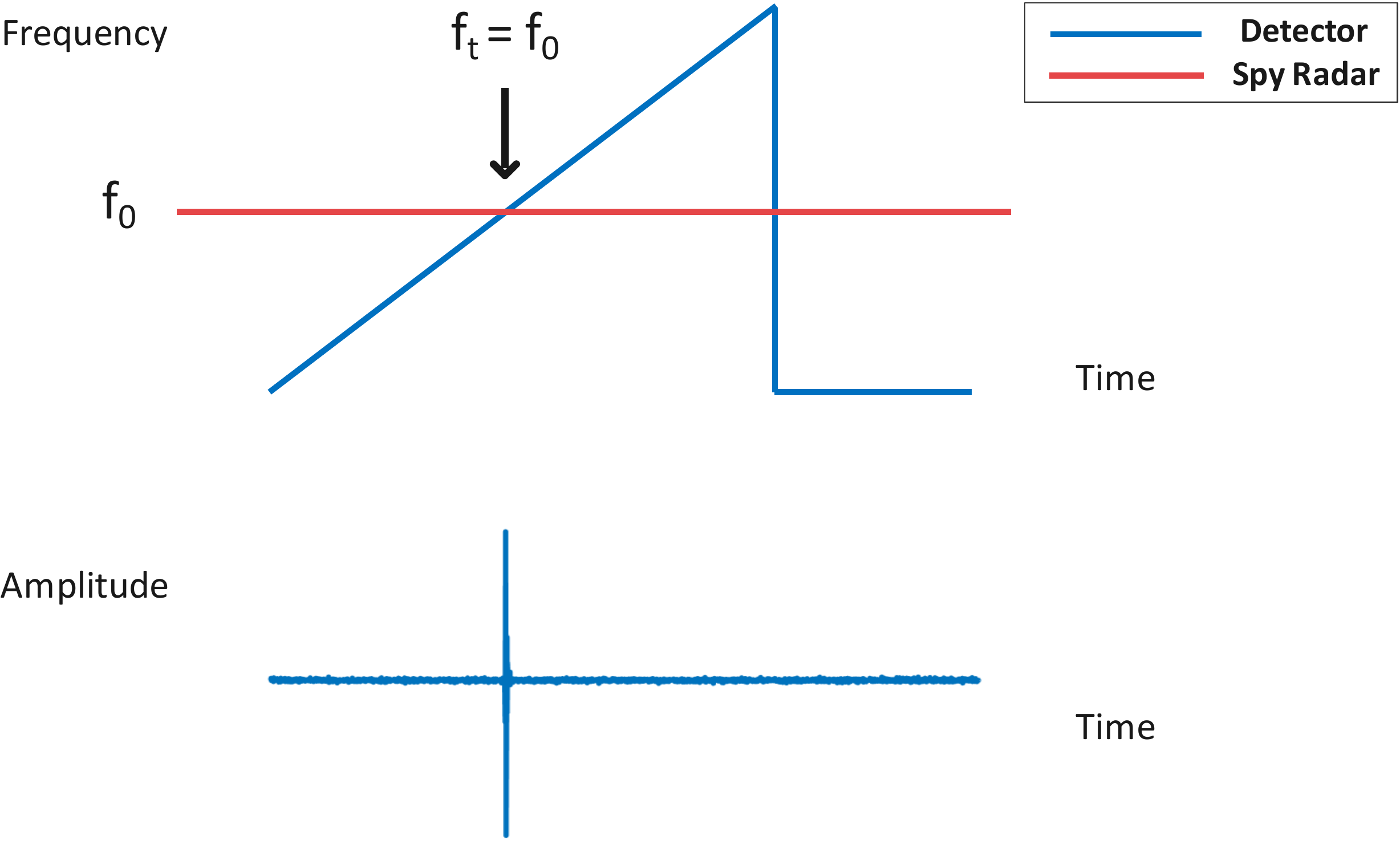}
    \caption{The observed peak in the time domain when $|f_t-f_0|<f_{cutoff}$.}
    \label{fig:Chirp Collision}
\end{figure}
After demodulation, the IF signal can be represented as 
\begin{equation}
    d(t)=r(t)\times c^*(t)=e^{j(2\pi (f_t-f_0)t)}.
\end{equation}

Only when $f_t$ is approaching $f_0$, or in other words, when $|f_t-f_0|<f_{cutoff}$, where $f_{cutoff}$ is the cutoff frequency of the low pass filter, the received signal can be demodulated to the baseband and sampled by its low sampling rate ADC. Otherwise, the low pass filter will remove all frequency components, and there is nothing in the baseband. As shown in Fig. \ref{fig:Chirp Collision}, the baseband exhibits a peak in the time domain if the received signal and the sweep-frequency signal have the same frequency components. Based on this observation, we propose to find peaks in the time domain to detect the presence of a narrow band mmWave signal.

Unlike the frequency of a narrow band signal that is invariant or only changes in a small range, the frequency of a wide band signal changes with time dramatically. Therefore, it is not suitable to use another sweep frequency signal to demodulate these wideband signals. This is because we do not know the exact frequency range of the spy radar, and it is hard to synchronize them. This causes a low probability that they have the same frequency components at the same time. However, we can search for a single-frequency signal to detect its presence.

Similarly, when the detector and the spy radar have the same frequency components, we can observe peaks in the time domain. However, since the wideband signal may not use the full mmWave band, we have to traverse the mmWave band to search for a single-frequency signal whose frequency is within the frequency range of the wideband signal.

Specifically, the detector demodulates the received signal using different single-frequency signals. We start at the minimum frequency $f_{min}$ of the mmWave band and set the frequency of the carrier $f_0=f_{min}$. By adding a suitable step size $\Delta f_0$ to increase $f_0$, the detector traverses the whole mmWave band. If no peak is found, it increases $f_0$ by $\Delta f_0$ until it reaches the maximum frequency $f_{max}$. There is a trade-off between search time and detection rate, and a smaller step size means a higher detection rate but a larger search time. To choose a suitable $\Delta f_0$, we discuss it in different scenarios.

For indoor applications, the mmWave radar usually adopts a large bandwidth since larger bandwidth means better range resolution \cite{TITutorial}: 
\begin{equation}
    \Delta d=c/2BW,
\end{equation}
where $c$ is the speed of light, and $BW$ is the bandwidth. 

However, the maximum range of the radar is limited by \cite{TITutorial}:
\begin{equation}
    d_{max}=\frac{f_s c}{2S},
\end{equation}
where $c$ is the speed of light, $f_s$ is the sampling rate, and $S$ is the frequency slope. For outdoor applications like autonomous vehicles, radar prefers a lower frequency slope. This is because it indicates a smaller bandwidth, to have a larger sensing range while sacrificing some range resolution.

Therefore, for indoor applications when the bandwidth is large, the detector prefers to use a large step size $\Delta f_0$ to reduce search time. For outdoor applications, a smaller step size is preferred for high detection rates.

\textit{Frequency Component Detection} is the most significant module in our design, and the following components are built based on this module.

\subsection{Waveform Classification}

After we detect the existence of the mmWave signal, we cannot directly conclude that a spy radar exists. This is because there are other mmWave devices also working in the mmWave band, such as WiGig transceivers. To differentiate these devices from spy radars, we designed a waveform classifier. As there is only one widely used mmWave band signal other than radar, which is WiGig, in this section, we only consider the classification between WiGig signals and radar signals. 

We first extract features that can characterize different signal patterns of radar and WiGig transceivers and then design a CNN-based waveform classifier for categorizing the signals.

\subsubsection{Feature Extraction}
\label{feaext}

\begin{figure}[t]
    \centering
    \includegraphics[width=250pt]{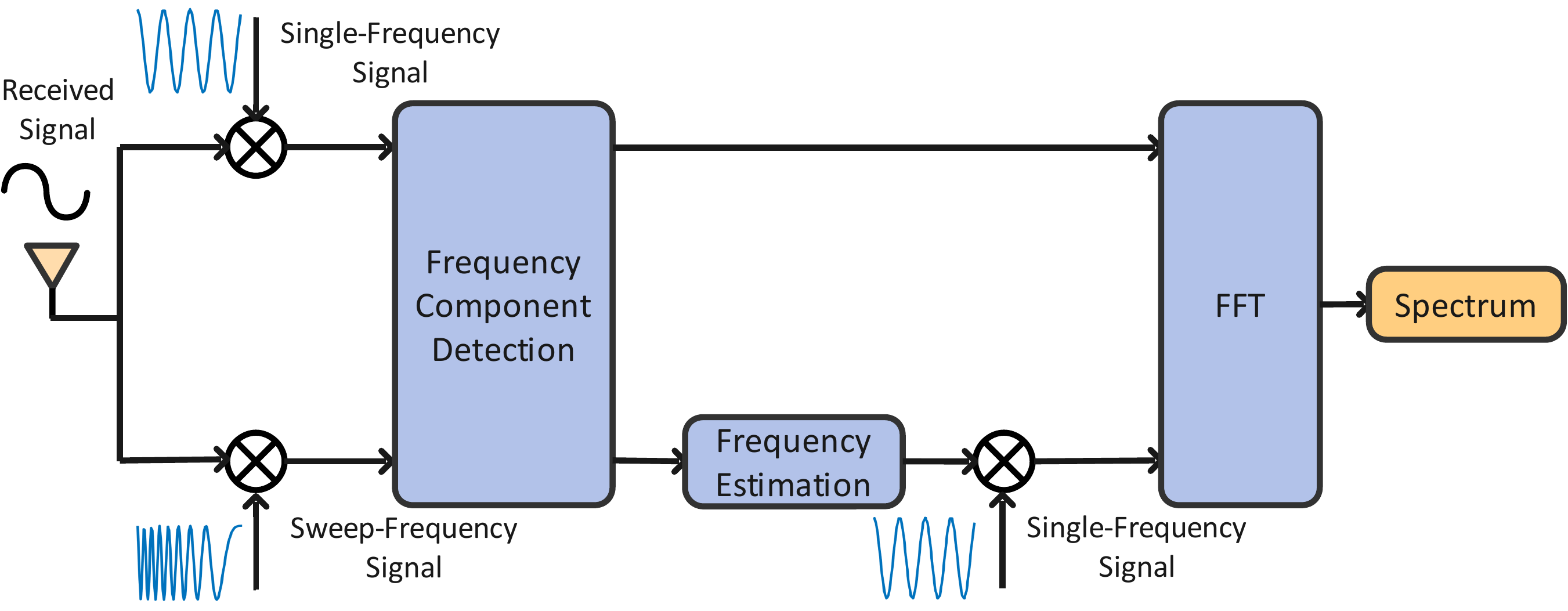}
    \caption{Feature Extraction. Spectrum features are extracted for waveform classification.}
    \label{fig:Feature Extraction}
\end{figure}

We observe that OFDM employed by WiGig has a different spectrum from radar signals \cite{OFDM}, which can be used for differentiating WiGig from radar signals.

To explore unique features for waveform classification, we will prove that single-frequency demodulation will not influence the shape of the spectrum, and thus the spectrum after single-frequency demodulation can be used for waveform classification.

Assume that the received signal has a frequency component of $f_t$, and we use a single-frequency signal with $f_0$ to demodulate it in the baseband. The demodulated signal can be written as
\begin{equation}
    d(t)=e^{j(2\pi (f_t-f_0)t)}.
\end{equation}

In the frequency domain, all frequency components are reduced by $f_0$, preserving the shape of the spectrum. However, demodulated with a sweep frequency signal, the spectrum will be changed since different frequency components are reduced to different values.

Based on this observation, we propose our feature extraction processing flow shown in Fig. \ref{fig:Feature Extraction}. If a wide band signal is received, we can simply get its spectrum by Fast Fourier Transformation (FFT) on the IF signal. While the case is more tricky for receiving a narrow band signal since the frequency of the narrow band signal is unknown to us. This brings challenges to demodulating it to the baseband and calculating the spectrum. Fortunately, after sweep-frequency signal demodulation, we are able to observe a peak when $|f_t-f_0|<f_{cutoff}$. As the frequency components of the sweep frequency can be estimated by its frequency slope and the time stamp. The frequency point of the narrow band signal $f_0$ is able to be estimated. Once we estimate the frequency point of the narrow band signal, we generate a single-frequency signal with the same frequency $f_0$, and its spectrum can be calculated using FFT.

\begin{figure}[t]
\centering
\subfigure[WiGig Spectrum]{
\begin{minipage}[t]{0.5\linewidth}
\centering
\includegraphics[width=1.5in]{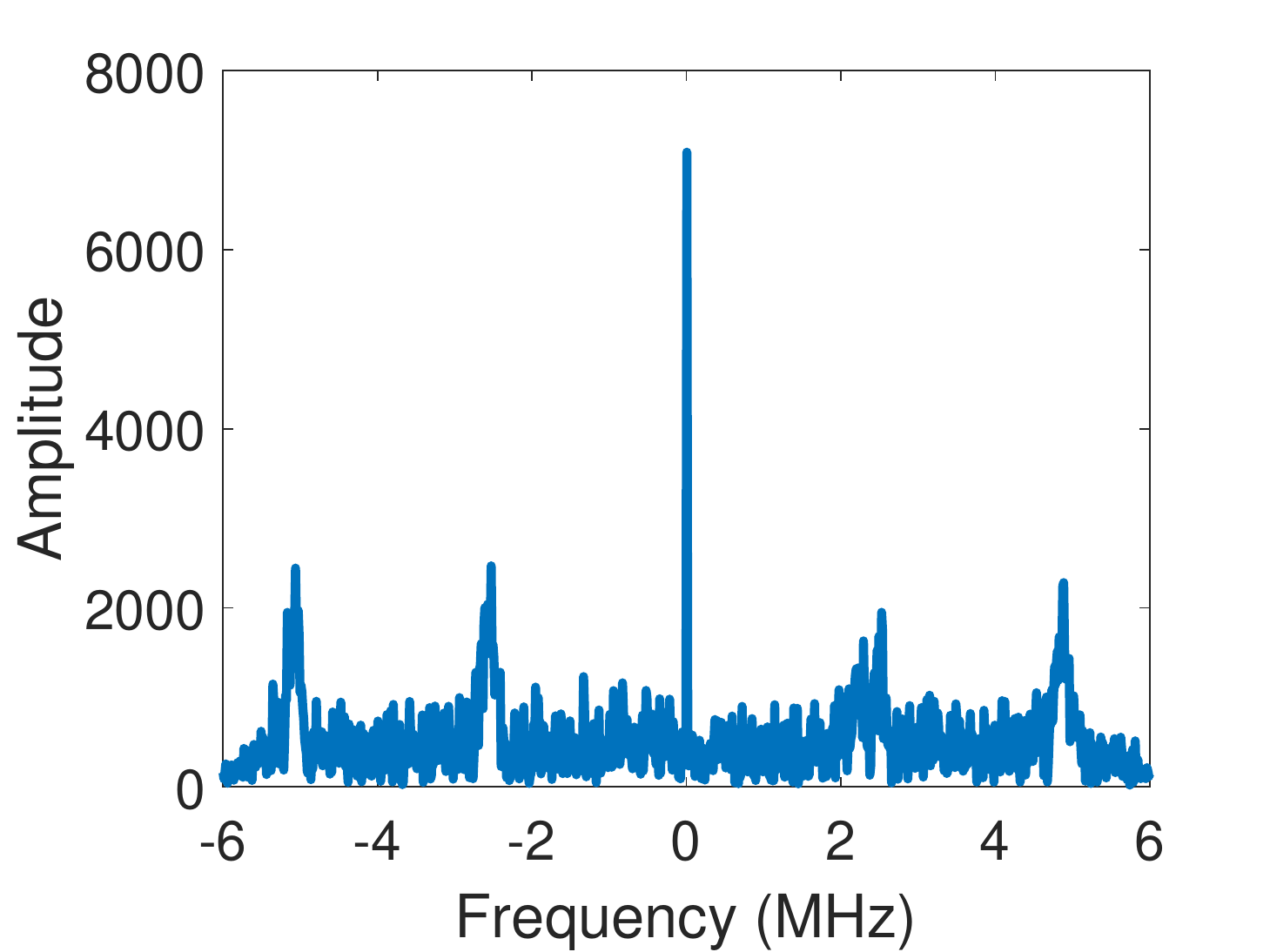}
\end{minipage}%
}%
\subfigure[CW/FSK Spectrum]{
\begin{minipage}[t]{0.5\linewidth}
\centering
\includegraphics[width=1.5in]{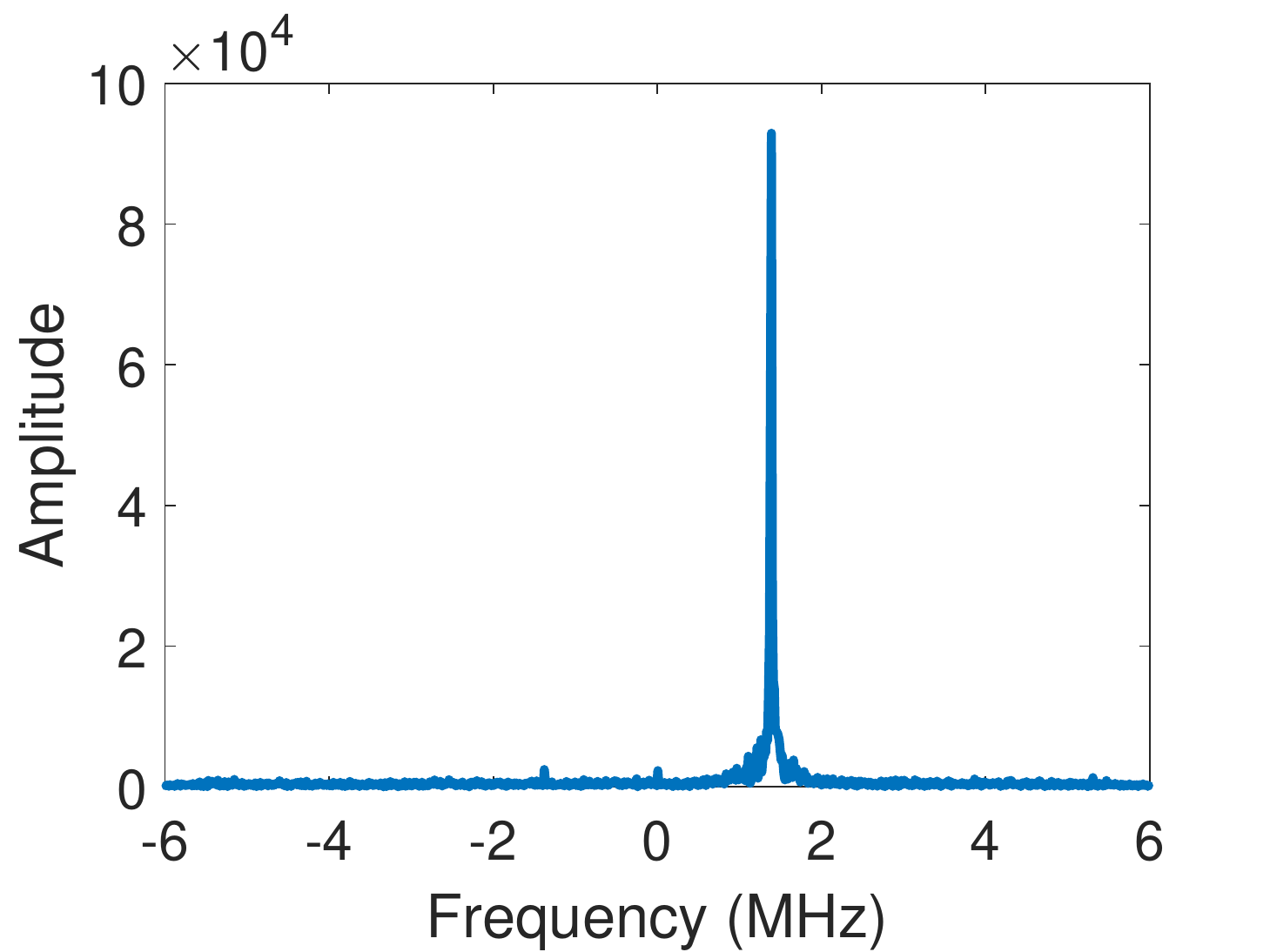}
\end{minipage}%
}%
\\
\subfigure[FMCW Spectrum]{
\begin{minipage}[t]{0.5\linewidth}
\centering
\includegraphics[width=1.5in]{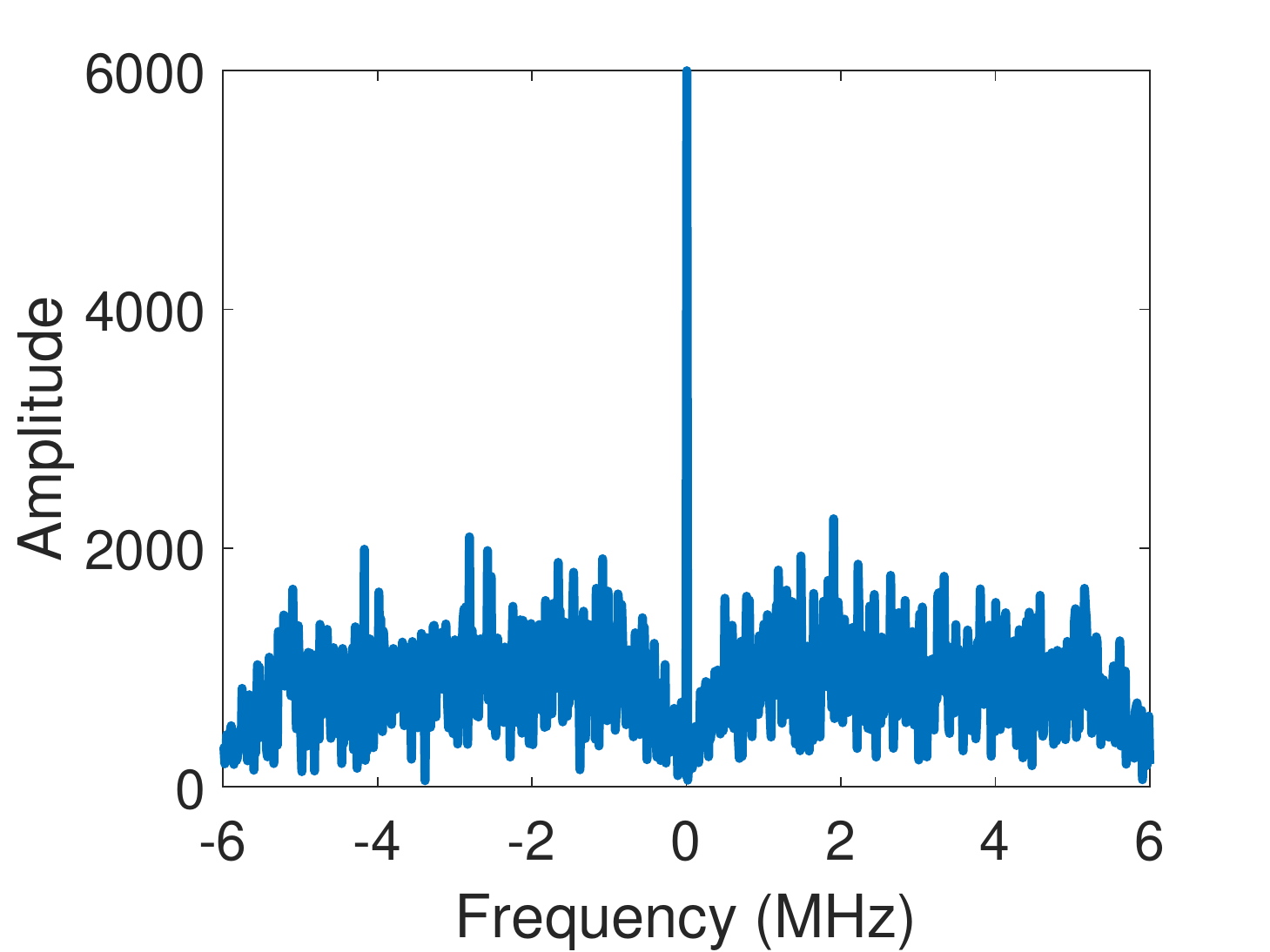}
\end{minipage}%
}%
\subfigure[Pulse radar Spectrum]{
\begin{minipage}[t]{0.5\linewidth}
\centering
\includegraphics[width=1.5in]{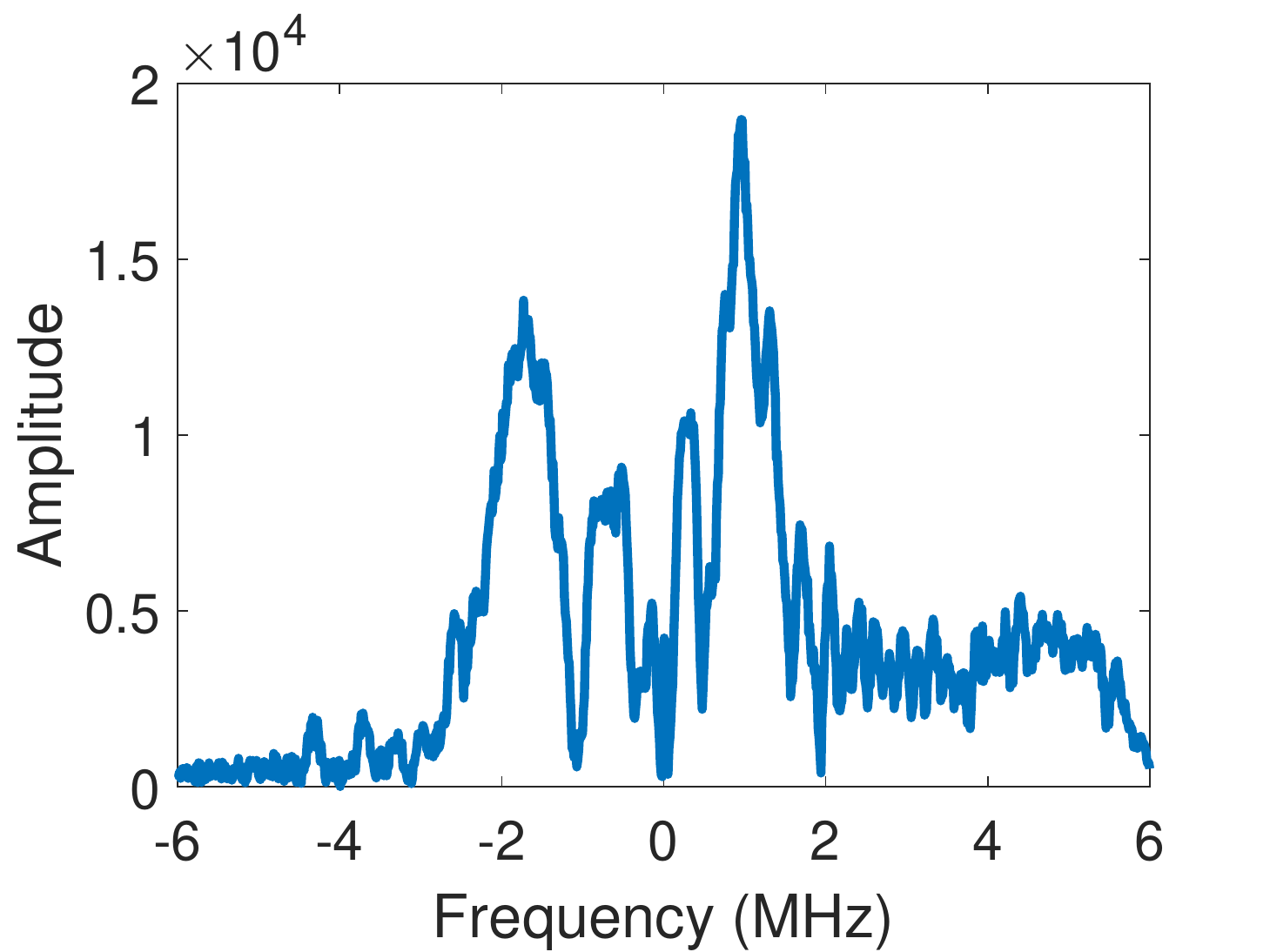}
\end{minipage}%
}%
\centering
\caption{Spectrum of various devices after single-frequency signal demodulation.}
\label{Fig: SpectrumCW}
\end{figure}

The spectrum of WiGig and radar signals is shown in Fig. \ref{Fig: SpectrumCW}. It reveals a large difference between WiGig and radar signals, which makes it possible to distinguish between them.

\subsubsection{CNN-based Classifier}

\begin{figure}[t]
    \centering
    \includegraphics[width=200pt]{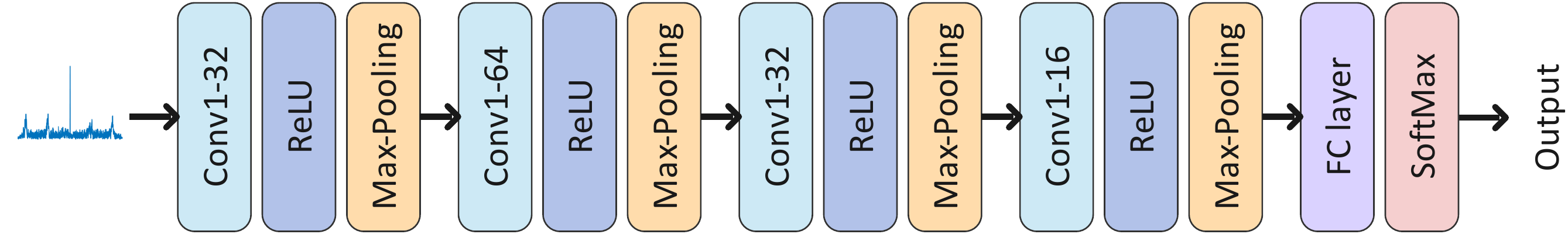}
    \caption{CNN-based Waveform Classifier }
    \label{fig:CNN}
\end{figure}

$Radar^2$ aims to recognize patterns of WiGig and radar on the spectrum. As shown in Fig. \ref{fig:CNN}, the input of our network is the 1$\times$1024 spectrum. The CNN architecture used in $Radar^2$ is a deep CNN consisting of four convolutional layers, each paired with a ReLU and a max pooling layer, followed by a fully connected (FC) layer. These convolutional layers encode the spectrum, producing a representation of the spectrum, and the FC layer outputs the probability of the type of signal. SoftMax is used to predict whether the signal comes from WiGig or radar.

\subsection{Spy Radar Detection}

A spy radar can be detected if (1) a signal in the mmWave band is detected, and (2) the signal can be classified as a radar signal. The whole process for spy radar detection is described in Algorithm \ref{alg:det}.

The detection result $q$ is initialized as ``false'' at first (line 1). To detect the existence of a mmWave band signal, a narrow band signal and a wide band signal should be detected respectively.

To detect a narrow band signal, we generate a sweep frequency signal $s(t)$ with the following parameters: amplitude $A_s$, start frequency $f_0$, frequency slope $S$, and initial phase $\phi_0$. The sweep frequency signal is
\begin{equation}
   s(t)=A_se^{2\pi(f_0+St)t+\phi_0}.
\end{equation}
The received signal $r(t)$ is mixed with $s(t)$, and the IF signal $y_s(t)$ is used for mmWave signal detection, $y_s(t)=r(t)s^*(t)$ (line 2).

To detect a wide band signal such as an FMCW signal, $Radar^2$ generates multiple single-frequency signals $l_i(t), i = 1, 2, 3, \cdots$:
\begin{equation}
   l_i(t)=A_ie^{2\pi f_it+\phi_i}, i = 1, 2, 3, \cdots
\end{equation}
where $A_i$ is the amplitude, $f_i$ is the carrier frequency and $\phi_i$ is the initial frequency of each single frequency signal. These single frequency signals mix with the received signal $r(t)$ respectively, and the IF signal $y_i(t), i = 1, 2, 3, \cdots$ is used for mmWave signal detection, $y_i(t) = r(t)l_i^*(t), i = 1, 2, 3, \cdots$ (line 3). The number of single frequency signals $i$ depends on the step size $\Delta f_0$. To choose a suitable step size $\Delta f_0$, we should consider the practical application, which we have discussed in \ref{fcd}.

We detect the existence of a signal in the mmWave band if a narrow band or wide band signal is identified as mmWave band signal. Whether there exists a signal in the mmWave band depends on the power of the IF signal $P(y(t))$ ($y(t)$ can be either $y_s(t)$ or $y_i(t), i = 1, 2, 3, \cdots$): if (1) the power is relatively large: $\frac{max(P(y_s(t)))}{mean(P(y_s(t)))}\geq R_{th}$, where $R_{th}$ is a ratio threshold, or (2) the power is sufficiently large: $max(P(y_s(t)))\geq P_{th}$, where $P_{th}$ is a power threshold, we detect the existence of a signal in the mmWave band (line 4 and line 14).

\begin{algorithm}[t]
    \caption{Spy Radar Detection}
    \label{alg:det}
    \begin{algorithmic}[1]
    \REQUIRE Received signal $r(t)$, a sweep frequency signal $s(t)$ with frequency slope $S$ and start frequency $f_0$, multiple single frequency signals $l_i(t), i = 1, 2, 3, \cdots$.\\
    \ENSURE Detection result: $q$.
    \STATE $q \leftarrow false$
    \STATE $y_s(t) \leftarrow r(t)s^*(t)$.\\
    \STATE $y_i(t) \leftarrow r(t)l_i^*(t), i = 1, 2, 3, \cdots$.\\
    \IF{$\frac{max(P(y_s(t)))}{mean(P(y_s(t)))}\geq R_{th}$ or $max(P(y_s(t)))\geq P_{th}$}
    \STATE $t_{max} \leftarrow \max_t P(y_s(t)) $
    \STATE $f_m \leftarrow f_0+S\cdot t_{max}$
    \STATE $y_m(t)\leftarrow r(t)l_m^*(t)$
    \STATE $F_m \leftarrow FFT(y_m(t))$
    \IF{$C(F_m)\neq WiGig$}
    \STATE $q \leftarrow true$
    \ENDIF
    \ENDIF
    \FORALL{$y_i(t), i = 1, 2, 3, \cdots$}
    \IF{{$\frac{max(P(y_i(t)))}{mean(P(y_i(t)))}\geq R_{th}$ or $max(P(y_i(t)))\geq P_{th}$}}
    \STATE $F_i \leftarrow FFT(y_i(t))$
    \IF{$C(F_i)\neq WiGig$}
    \STATE $q \leftarrow true$
    \ENDIF
    \ENDIF
    \ENDFOR
    \RETURN $q$
    \end{algorithmic}
\end{algorithm}

Once a mmWave signal is detected, its spectrum is further calculated for signal classification. If it is not WiGig, $Radar^2$ identifies the existence of a spy radar. In Section \ref{feaext}, we have demonstrated that single-frequency signal demodulation will not affect the original spectrum, but sweep-frequency signal demodulation will destroy its original spectrum. To obtain the spectrum of a narrow band signal, $F_i$, we perform FFT on $y_i(t), i = 1, 2, 3, \cdots$ (line 15). But $y_s(t)$ cannot be used for spectrum analysis because it is demodulated by a sweep frequency signal. $Radar^2$ should estimate a frequency component $f_m$ for the wideband signal and generate a single frequency signal for spectrum analysis. To estimate the frequency component $f_m$, we leverage the frequency component found in $y_s(t)$. Specifically, once the time when the power is maximum, $t_{max}$, has been detected, the frequency component $f_m$ can be calculated by: $f_m=f_0+S\cdot t_{max}$ (line 5-6), where $f_0$ is the start frequency, and $S$ is the frequency slope of the sweep frequency signal. We generate a single frequency signal $l_m(t)$ using this frequency component $f_m$:
\begin{equation}
    l_m(t)=A_me^{2\pi f_mt+\phi_m}.
\end{equation}
$l_m(t)$ is mixed with $r(t)$ again to get the wide band signal spectrum $F_m$ (line 7-8).

Finally, the spectrum $F_i$ or $F_m$ is fed into the waveform classifier $C(\cdot)$. If the signal is not recognized as $WiGig$, $Radar^2$ identifies the existence of a spy radar, and ``true'' is assigned to the detection result $q$ (lines 9-11 and lines 16-18).

\subsection{Localization}

To localize spy radars, $Radar^2$ cannot use the conventional wireless localization method, because it cannot differentiate environmental objects from spy radars. The main difference between spy radars and objects is that the objects do not transmit any signal while spy radars are active. We exploit this feature to localize spy radars. Specifically, we move our detector to multiple known positions. At each position, we estimate the direction of a spy radar using the MUSIC algorithm. \textit{Triangulation} is used to locate the radar spy. Localizing spy radars can guide users to take further action, like removing them.

\subsubsection{AoA Estimation}

\begin{figure}[t]
    \centering
    \includegraphics[width=150pt]{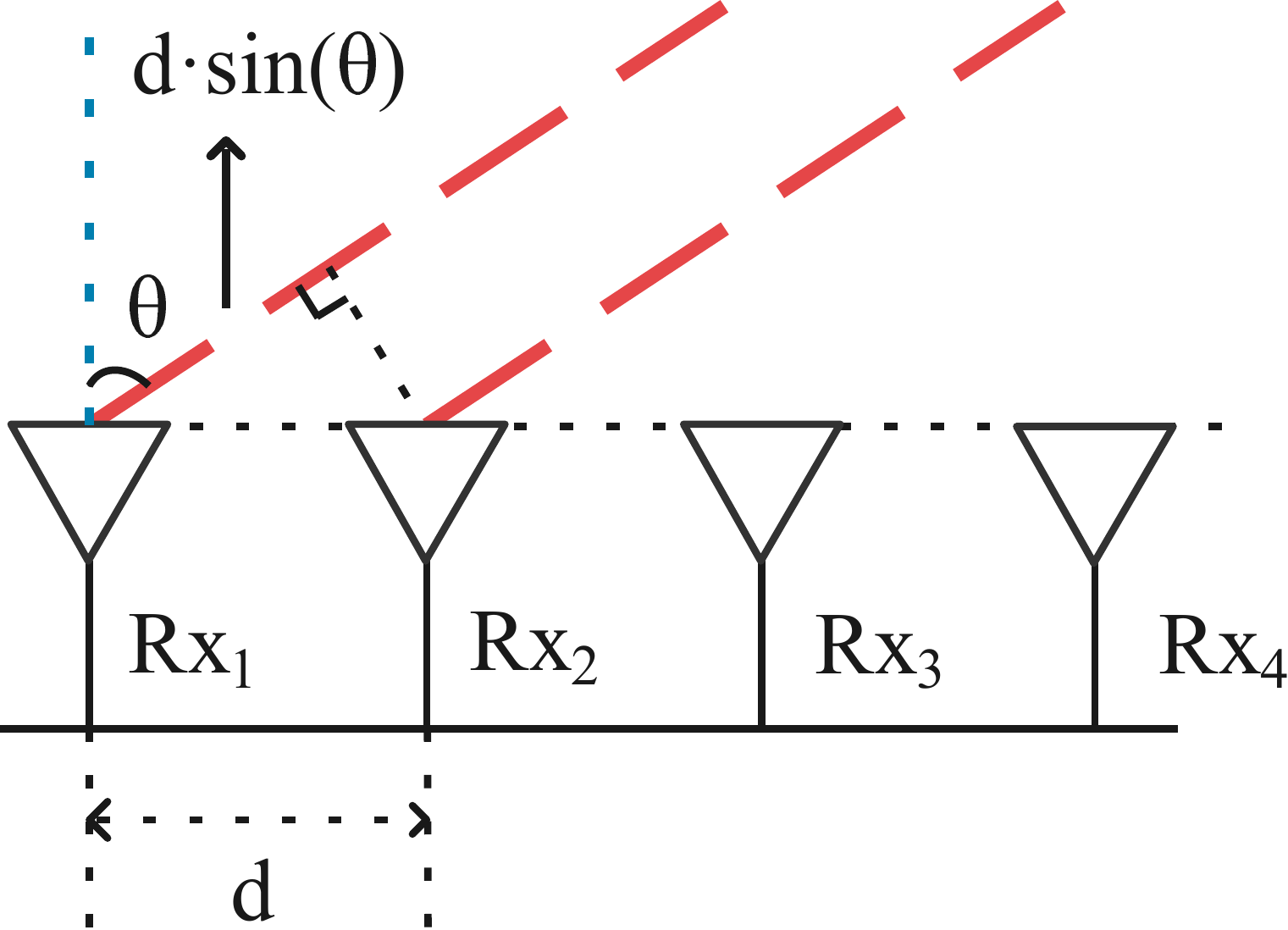}
    \caption{AoA estimation}
    \label{fig:AoAestimation}
\end{figure}

Triangulation is based on multiple observations of Angle-of-Arrival (AoA). We illustrate the basic idea of AoA estimation in Fig. \ref{fig:AoAestimation}. The objective is to estimate the AoA $\theta$. The distance between different receiving antennas $d$ is known. Based on the geometric relationship, the wave path difference between different receiving antennas is $d \cdot sin(\theta)$. Suppose the received signal from the first antenna is
\begin{equation}
    r_1(t)=A_1e^{2\pi f_t t+\phi_0},
\end{equation} 
then the received signal from the second antenna is
\begin{equation}
    r_2(t)=A_2e^{2\pi f_t (t+\frac{2d \cdot \sin(\theta)}{c})+\phi_0}.
\end{equation}
The phase difference between $r_1(t)$ and $r_2(t)$ is 
\begin{equation}
    \Delta \phi = \frac{4\pi f_t d \cdot \sin(\theta)}{c}=\frac{4\pi d \cdot \sin(\theta)}{\lambda}.
\end{equation} 

Therefore, the AoA $\theta$ could be estimated from the phase difference of different receiving antennas as
\begin{equation}
    \theta=\arcsin(\frac{\lambda\Delta \phi}{4\pi d}).
\end{equation}

We use the MUSIC algorithm \cite{kundu1996modified, engholm2011direction, zhang2004doa, gupta2015music} to estimate AoA at each anchor. Once $\theta_i$ is estimated at different anchors, it can estimate the location of the spy radar by triangulation.

\subsubsection{Triangulation}
\label{triangulation}
\begin{figure}[t]
    \centering
    \includegraphics[width=150pt]{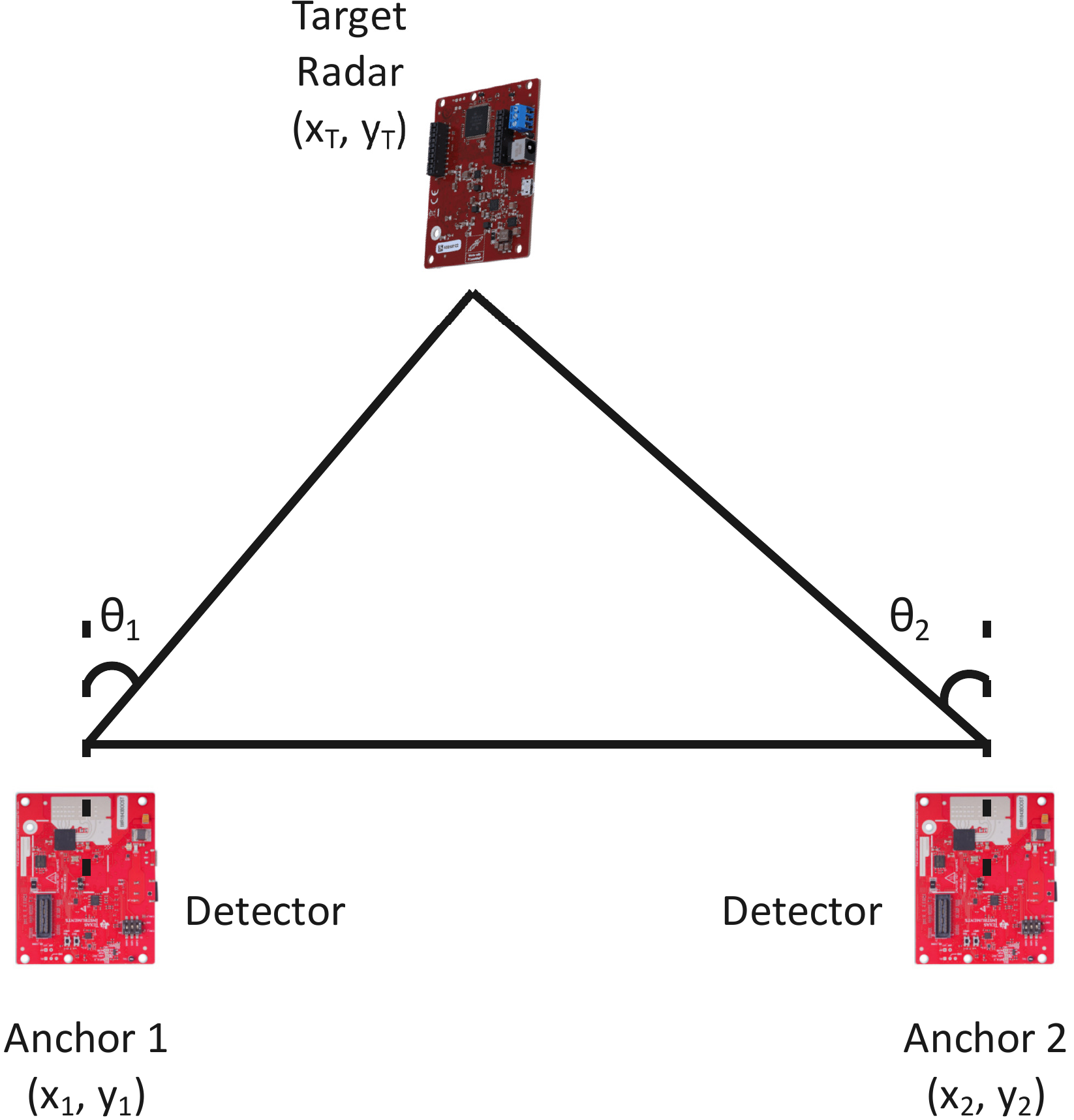}
    \caption{Triangulation}
    \label{fig:TriLozalication}
\end{figure}

We move the detector to multiple known positions (anchors) and estimate the AoA of the spy radar at each anchor. Triangulation can be used to estimate the position of the spy radar \cite{maxim2008Triangular, oguejiofor2013Triangular}. 

Given $N$ anchor points, for each anchor $i$, $i=1, 2, \cdots, N$, the estimated AoA is $\theta_i$, $Radar^2$ draws a line through anchor $i$ along the direction $\theta_i$. Ideally, the intersection of the above-mentioned $N$ lines should be the location of the spy radar. Due to AoA estimation errors, the lines cannot intersect at one point. The problem of spy radar localization can be converted to the problem of finding the optimal point, which has the shortest sum distance to all lines. We formulate the problem mathematically in Appendix and present an SVD-based solution to find the position of the spy radar \cite{han2010nearest}.

\subsection{Multi-Device Scenario}
\label{multidev}

$Radar^2$ can also work well in multi-device scenarios, where multiple spy radars or WiGig devices coexist. The system design described in previous subsections is a basic version for single-device detection. In this subsection, we will describe the modification of each module to support multi-device detection and localization.

\subsubsection{Frequency Component Detection}

This module is designed to detect the presence of mmWave signals. It can detect all frequency components of signals in the mmWave band irrespective of whether it is a single device or multiple devices. So this module remains unchanged.

\subsubsection{Waveform Classification}

Waveform Classification is designed to differentiate WiGig from radar signals. If the detector receives both radar signals and WiGig simultaneously, it is important to separate them and identify them correctly. In other words, $Radar^2$ should be able to detect the presence of a mmWave radar when there exists another radar or another WiGig transceiver.

Before waveform classification, we need to add a function \textit{Signal Separation} for the multi-device scenario. We assume that different devices work at different positions, so their AoA is unique, and we can use MUSIC to determine the direction of each signal. Thus we can separate signals sent by different devices. Specifically, we first estimate the number of devices by counting the number of peaks in AoA estimation. Then we separate them by their direction and feed them to our waveform classifier separately.

\subsubsection{Spy Radar Detection}

This module combines the results of \textit{Frequency Component Detection} and \textit{Waveform Classification}. The logic of the decision-maker module only needs a small change for the multi-device scenario: if mmWave signals are detected, and one of them can be identified as a radar signal, we detect the presence of spy radar. Besides, this module outputs another parameter, the number of spy radars $n_{i}$, to the next module \textit{Localization} in a multi-device scenario. The number of spy radars depends on the number of signals identified as radar signals in the module \textit{Waveform Classification}.

\subsubsection{Localization}

Since we have known the number of spy radars from \textit{Spy Radar Detection} module, and it can estimate their direction using MUSIC, we can localize them one by one using \textit{Localization} module. Notice that in this module, only AoA identified as spy radar will be considered for localization.

When multiple spy radars exist, we should move to three or more anchors to locate the positions of spy radars \cite{sogo2000real}.

In this case, because sometimes a spy radar can be blocked by other spy radars, the detector may observe a different number of spy radar $n_{i}, i=1, 2, \cdots, N$ at each anchor, where $N$ is the number of anchors. However, we assume that among all anchor points, at least one can observe all spy radars. So we choose the maximum of $n_i$ as the number of spy radars:
\begin{equation}
    \hat{n}=\max(n_i),i=1, 2, \cdots, N.
\end{equation}
The multiple radars localization problem is solved based on single spy radar localization. $Radar^2$ chooses one AoA at each anchor as if there is only one spy radar, then the problem is reduced to single radar localization. By solving the problem, we can calculate a position and its corresponding error. By traversing all combinations of AoA, we can get all possible radar locations. Among these locations, we choose $\hat{n}$ locations with the smallest errors.

Based on this design, our system can detect and localize spy radars when the WiGig transceiver and radars coexist.

\section{Implementation and Evaluation}
\label{evaluation}

\begin{figure*}[t]
\centering
\begin{minipage}[t]{0.49\linewidth}
\centering
\subfigure[mmWave Radar testbed.]{
\begin{minipage}[t]{0.49\linewidth}
\centering
\includegraphics[width=1.28in]{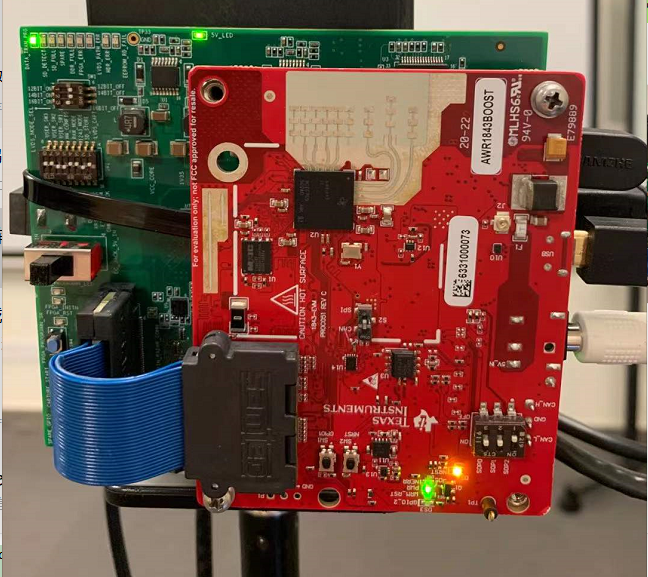}
\end{minipage}%
}%
\subfigure[Scenario]{
\begin{minipage}[t]{0.49\linewidth}
\centering
\includegraphics[width=1.5in]{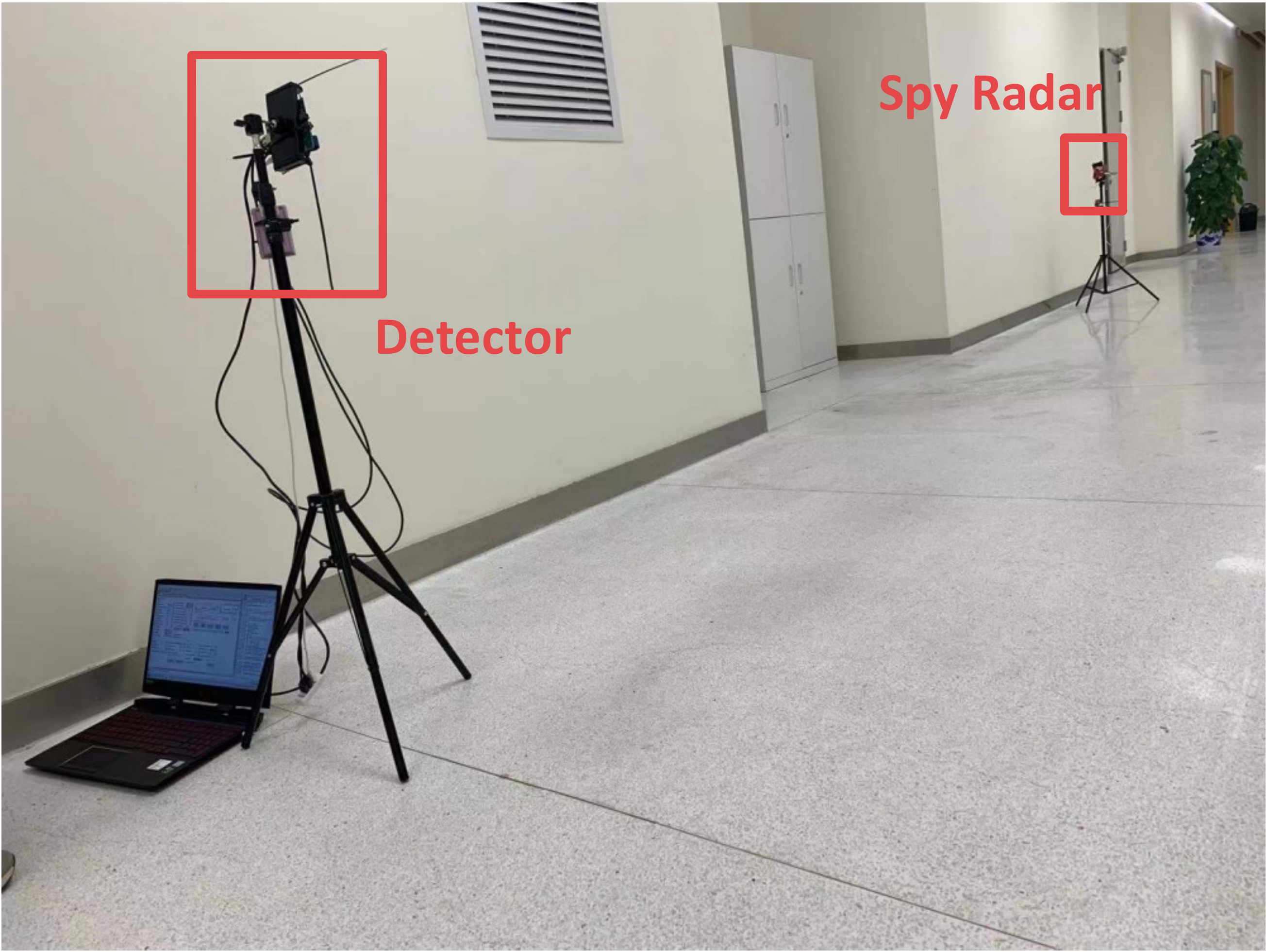}
\end{minipage}%
}%
\caption{Experiment Setup}
\label{fig: setup}
\end{minipage}
\begin{minipage}[t]{0.49\linewidth}
\centering
\subfigure[Room Layout 1: open area]{
\begin{minipage}[t]{0.49\linewidth}
\centering
\includegraphics[width=1.13 in]{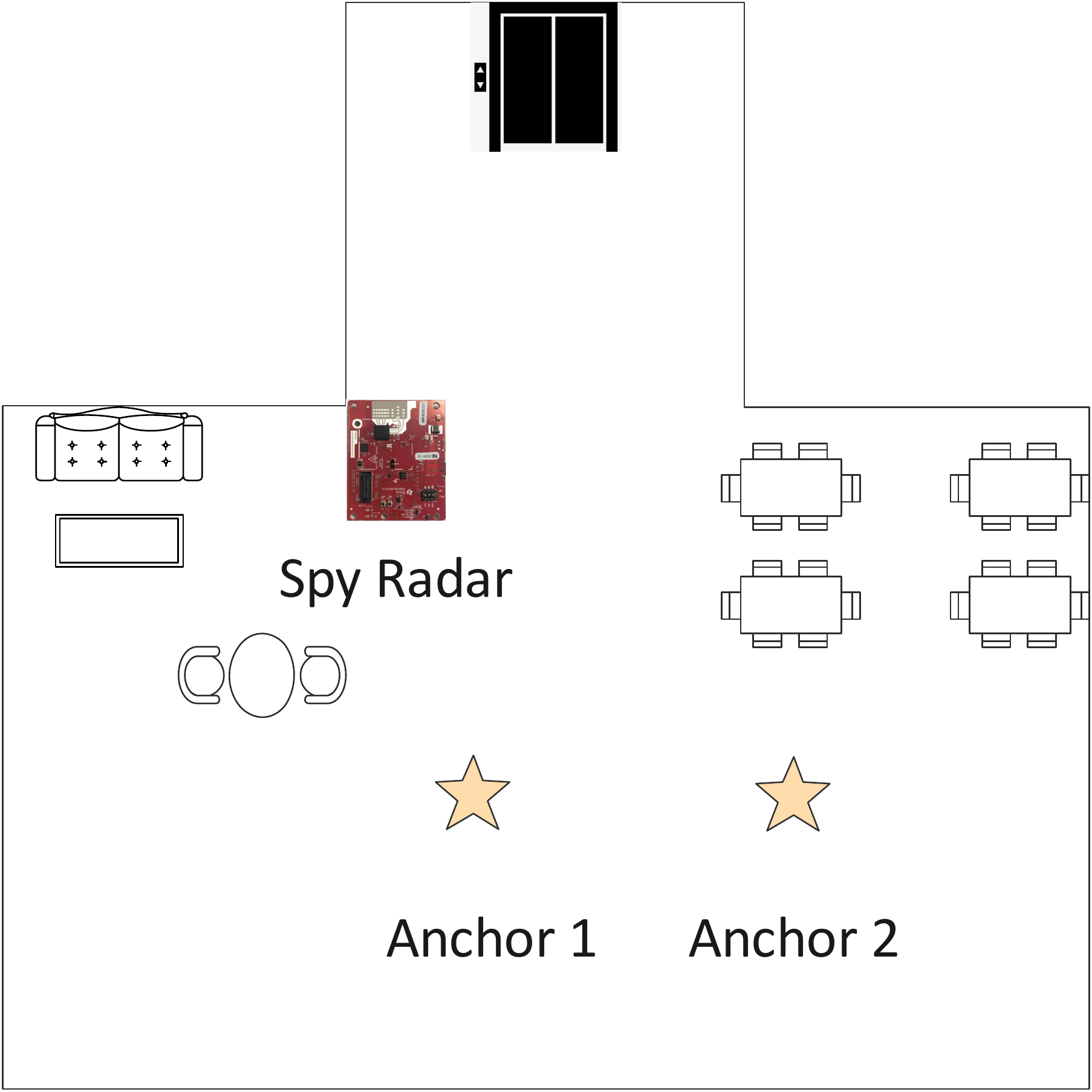}
\end{minipage}%
}%
\subfigure[Room Layout 2: a conference room]{
\begin{minipage}[t]{0.49\linewidth}
\centering
\includegraphics[width=1.13 in]{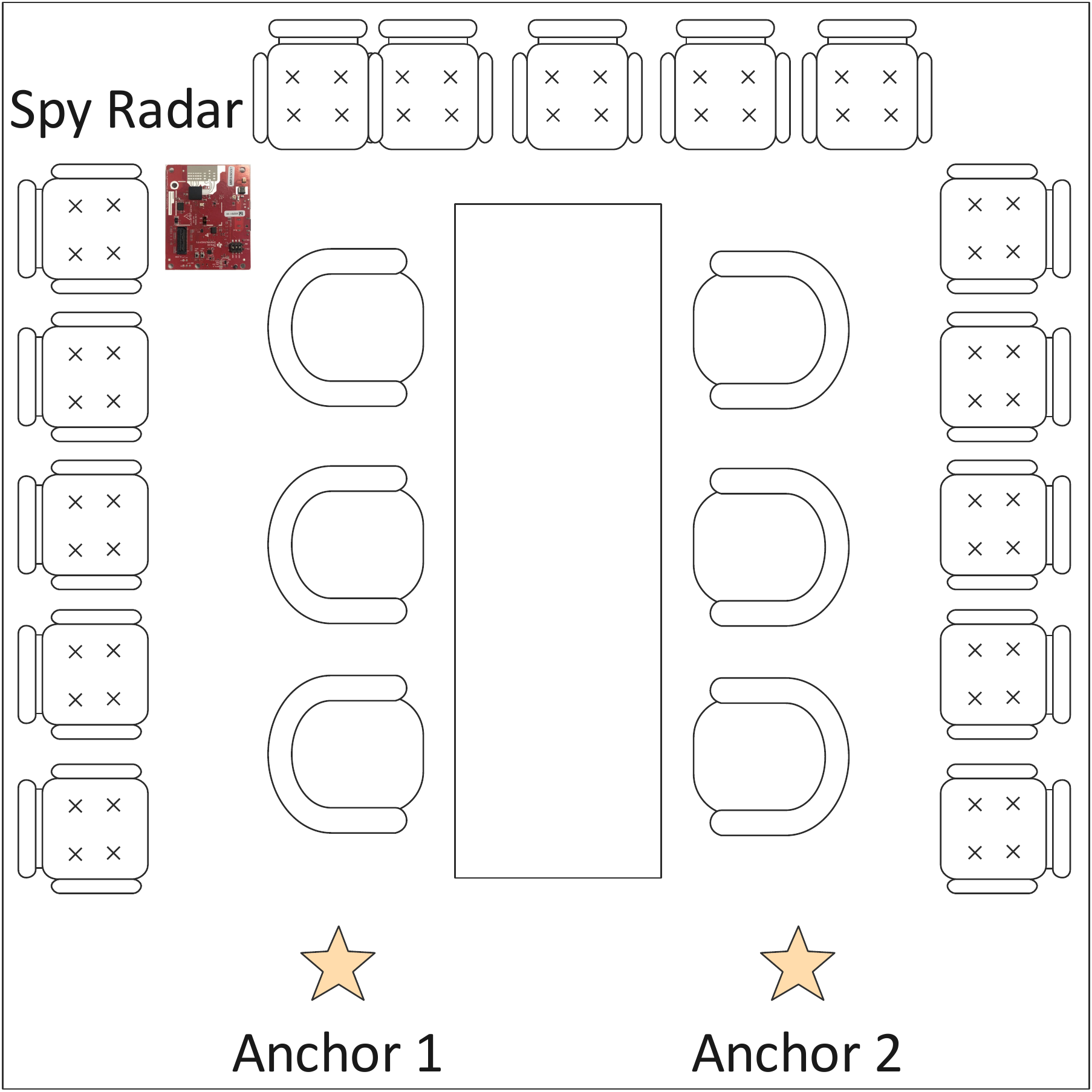}
\end{minipage}%
}%
\caption{Room Layout.}
\label{fig: roomlayout}
\end{minipage}
\end{figure*}

We implement $Radar^2$ using COTS mmWave radars. We use TI IWR1443 \cite{iwr1443} and TI AWR1843 \cite{awr1843} radars for 77-81GHz frequency band radar detection, and TI IWR6843 \cite{iwr6843} for 60-64GHz frequency band detection. 

In $Radar^2$, we set the parameters of the COTS radar as described in Table~\ref{tab:DetectorPara}. Four onboard receiving antennas are leveraged to estimate the AoA of the spy radar. To detect the presence of a single frequency signal, we designed a sweep frequency signal that covers the mmWave band of the testbed.

The frequency slope of the sweep frequency signal is set to 39.9756$MHz/ \mu s$. Similarly, to detect the presence of a sweep frequency signal, we design multiple single-frequency signals that traverse the whole mmWave band of the testbed. The step frequency $\Delta f_0$ of these single frequency signals is set to 800MHz.

$Radar^2$ uses 25 frames (about 0.8 seconds) signal to detect the presence of mmWave signals. As a result of our empirical evaluation, we choose the ratio threshold $R_{th}$ as 4.4, and the power threshold $P_{th}$ as 70 dBm. In addition, we perform 1024-point FFT and normalize the spectrum for waveform classification. Our CNN architecture is implemented on Pytorch 1.10.0 using Cuda 11.2.

\begin{table}[t]
    \centering
    \caption{Parameter setting.}
    \begin{tabular}{c | c}
    \hline
    Parameter & Value\\
    \hline
       Sweep Time (us) & 100\\
       Chirp Idle Time (us) & 10\\
       Sampling Rate (kbps) & 12000\\
       ADC samples & 1024\\
       Frame Period (ms) & 33.3\\
       Number of chirps in a frame & 128\\
    \hline
    \end{tabular}
    \label{tab:DetectorPara}
\end{table}

\subsection{Experiments}

We evaluate the $Radar^2$ in both single-device and multi-device scenarios, respectively.

\subsubsection{Single-Device Experiment}

We use IWR1443 and AWR1843 as spy radars on 77GHz, use IWR6843, Infineon BGT60LTR11-AIP \cite{Infineon} as spy radars on 60GHz, and use GuanYee WIHD-551 \cite{GuanYee} as the WiGig transceivers on 60GHz. The spy radar transmits a CW/FSK/FMCW/pulse signal in space for their sensing. For the $Radar^2$ detector, we disable its transmission antennas and only enable its receiving antennas for radar detection, which can be implemented by mmWave Studio software provided by TI \cite{mmWaveStudio}.

(1) We evaluate the performance of spy radar detection under various impact factors.  

\textbf{Environment.} We chose two different indoor environments and one outdoor environment for evaluation. The room layouts are shown in Fig. \ref{fig: roomlayout}. Layout1 is an open area where the multi-path is not severe, while Layout2 is a conference room with a severe multi-path effect. Two AWR1843 boards are deployed in this experiment; one works as a spy radar while the other works as a detector. A detector is placed within 15 degrees of the spy radar at a distance of five meters.

\begin{figure}[t]
    \centering
    \includegraphics[width=150pt]{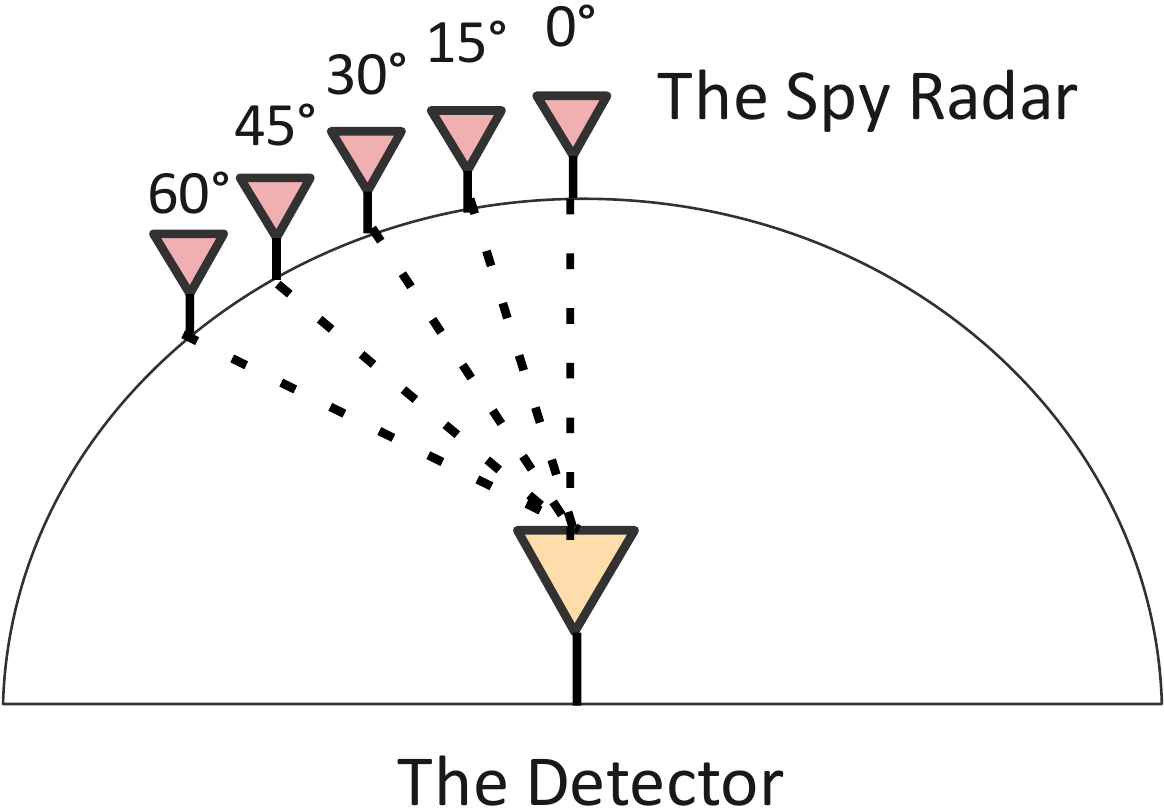}
    \caption{Angle Setup}
    \label{fig:orientation}
\end{figure}

\textbf{Distance and Angle.}
We placed our detector at different distances and angles in a very long corridor to explore the maximum sensing range of our system. First, we compare the spy radar detection rate when the detector is placed at a distance of 5 m, 10 m, 15 m, 20 m, and 25 m from the radar. Then, we move the detector 3m away from the spy radar and tune the angle of the spy radar within the Field of View (FoV) of the detector. As shown in Fig. \ref{fig:orientation}, since the FoV of our detector is 120$\degree$, we change the angle to be 0$\degree$, 15$\degree$, 30$\degree$, 45$\degree$, and 60$\degree$ respectively and evaluate the spy radar detection rate to investigate the impact of angle in our system.

\textbf{Testbeds $\&$ Types of Radar.}
We use different combinations of $\{$spy radar, detector$\}$ to evaluate the robustness of our system against different testbeds and different types of radar. TI IWR1443 and AWR1843 are used as 77GHz platforms, and IWR6843 and Infineon BGT60LTR11AIP are used as 60GHz platforms. A detector is placed within 15 degrees of the spy radar at a distance of three meters in a corridor, and we set the spy radar to work in different modes (CW/FMCW/FSK/Pulse). TI radars are able to send FMCW and FSK signals, while the Infineon BGT60LTR11AIP can work in two different modes: Doppler (CW) mode and pulse mode.

\textbf{Human Activity.}
To explore the impact of human activity, people are asked to walk around the spy radar and the detector. The person sometimes blocks the line-of-sight (LOS) channel between the detector and the spy radar. A detector is placed in a corridor within 15 degrees of the spy radar and at a distance of five meters.

(2) We also evaluated the performance of localization under different impact factors.

\textbf{Environment.} As described above, different environments have different severity of multi-path. To verify the robustness of localization against the environment, we placed the spy radar at different locations in three environments: two indoor environments, which are shown in Fig. \ref{fig: roomlayout}, and an outdoor environment.

\textbf{Height Difference.} To investigate the impact of height difference on 2D localization, we set the height difference of the spy radar and the detector to 0.33m, 0.66m, and 1m, respectively, and evaluate the performance of localization. The experiment is conducted in a corridor using an AWR1843 testbed. We move the detector to two anchors to measure the location of a spy radar, and we move the spy radar to five different positions for evaluation.

\textbf{Number of anchors.} Number of anchors has an impact on localization performance. We compare localization errors using 2-6 anchors to investigate the selection of the number of anchors. AWR1843 is used in this experiment.

\subsubsection{Multi-Device Experiment}

In this experiment, we investigate the performance of the system when multiple radars working in different modes (FMCW/CW/FSK/pulse) coexist, or a WiGig device and a spy radar coexist. 

We compare the detection rate and localization error with the single-device scenario. In this experiment, two devices are placed at 20$\degree$ and -20 $\degree$ of 5m from the detector.

\begin{figure*}[t]
\begin{minipage}[t]{0.49\linewidth}
\centering
\subfigure[Impact of Distance]{
\begin{minipage}[t]{0.5\linewidth}
\centering
\includegraphics[width=1.47 in]{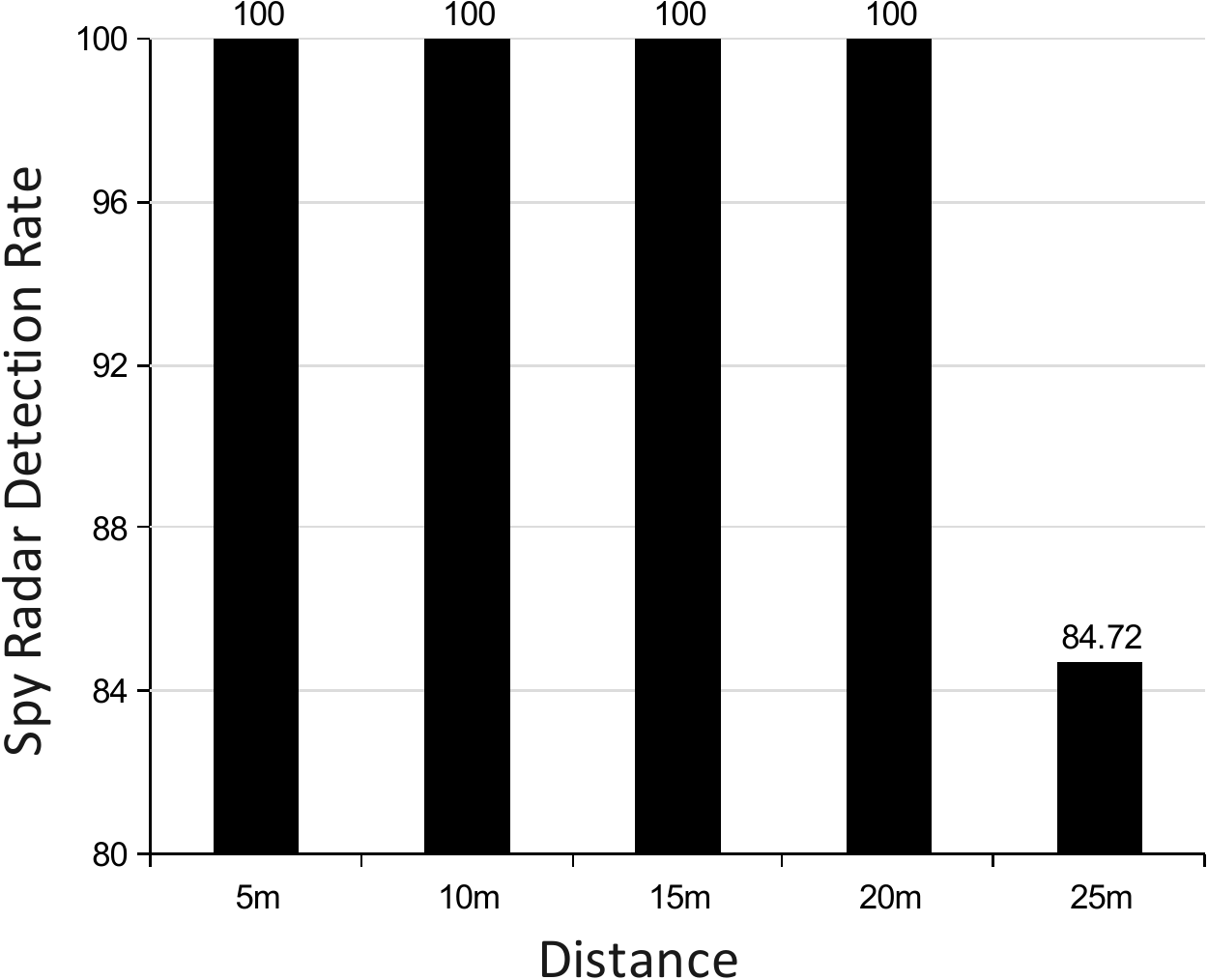}
\end{minipage}%
}%
\subfigure[Impact of Angle]{
\begin{minipage}[t]{0.5\linewidth}
\centering
\includegraphics[width=1.47 in]{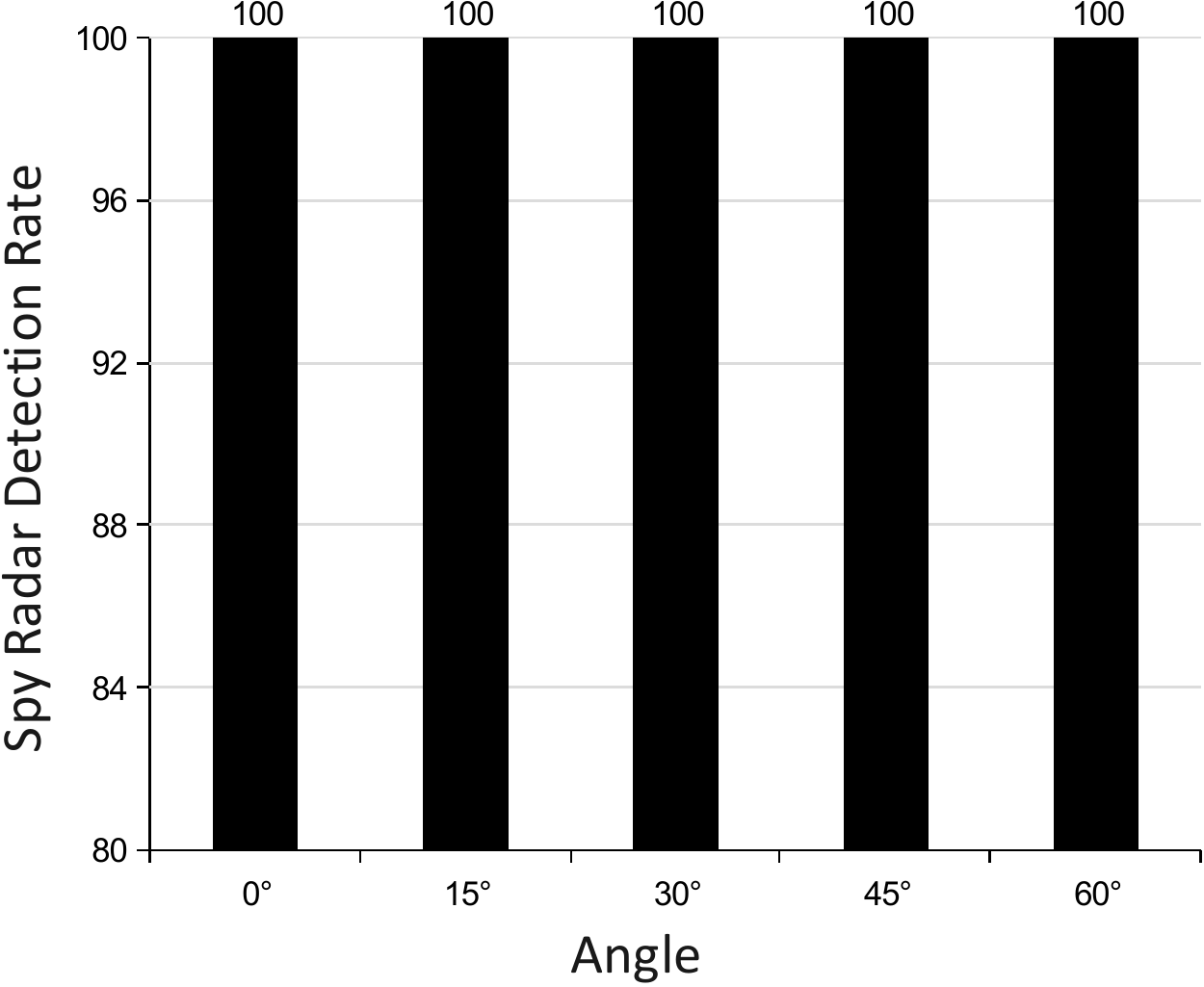}
\end{minipage}%
}%
\centering
\caption{Detection rate at various distances and angles.}
\label{fig: Detdisang}
\end{minipage}
\begin{minipage}[t]{0.49\linewidth}
\centering
\subfigure[Angle Error CDF]{
\begin{minipage}[t]{0.5\linewidth}
\centering
\includegraphics[width=1.6 in]{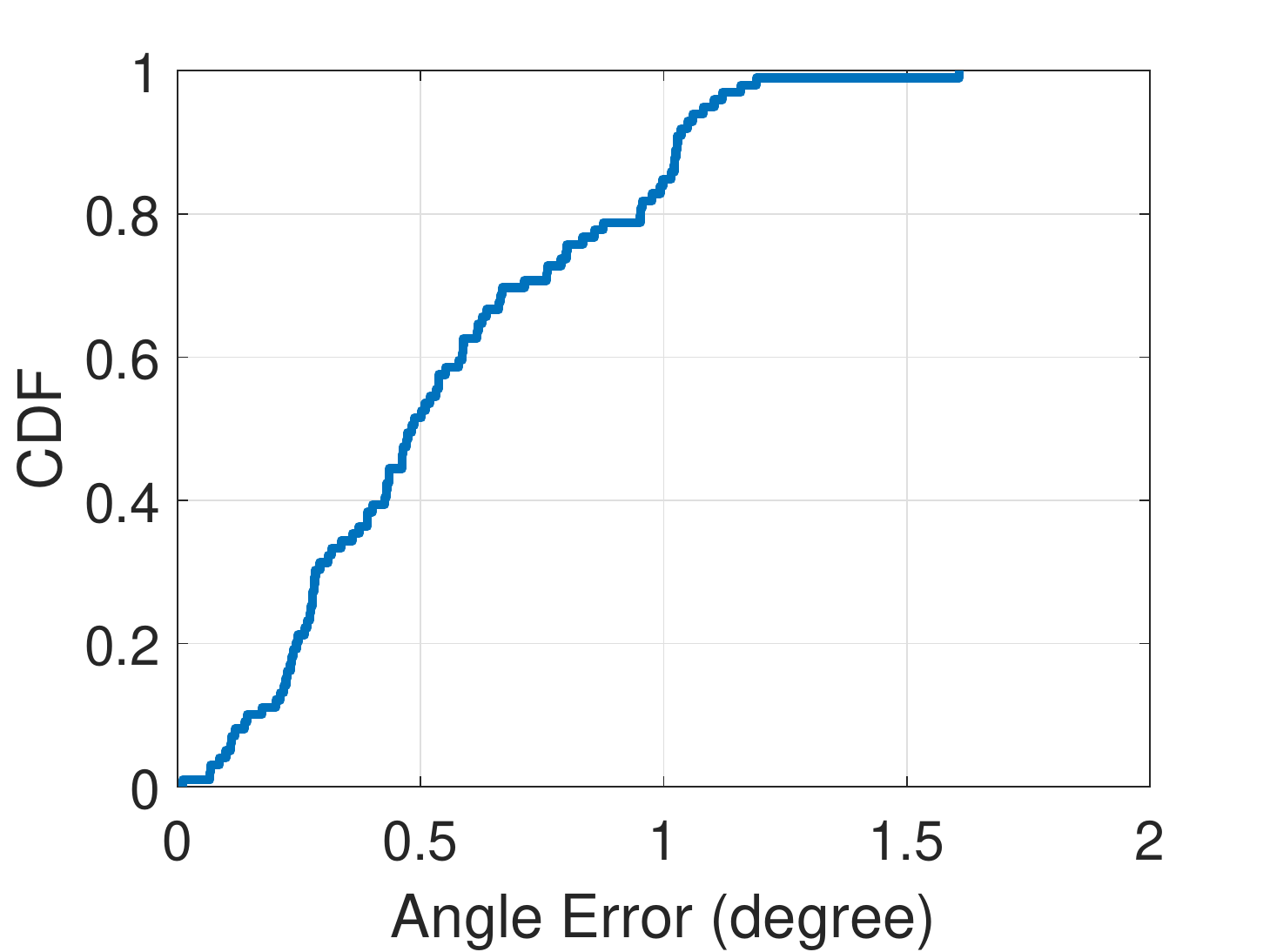}
\end{minipage}%
}%
\subfigure[Localization Error CDF]{
\begin{minipage}[t]{0.5\linewidth}
\centering
\includegraphics[width=1.6 in]{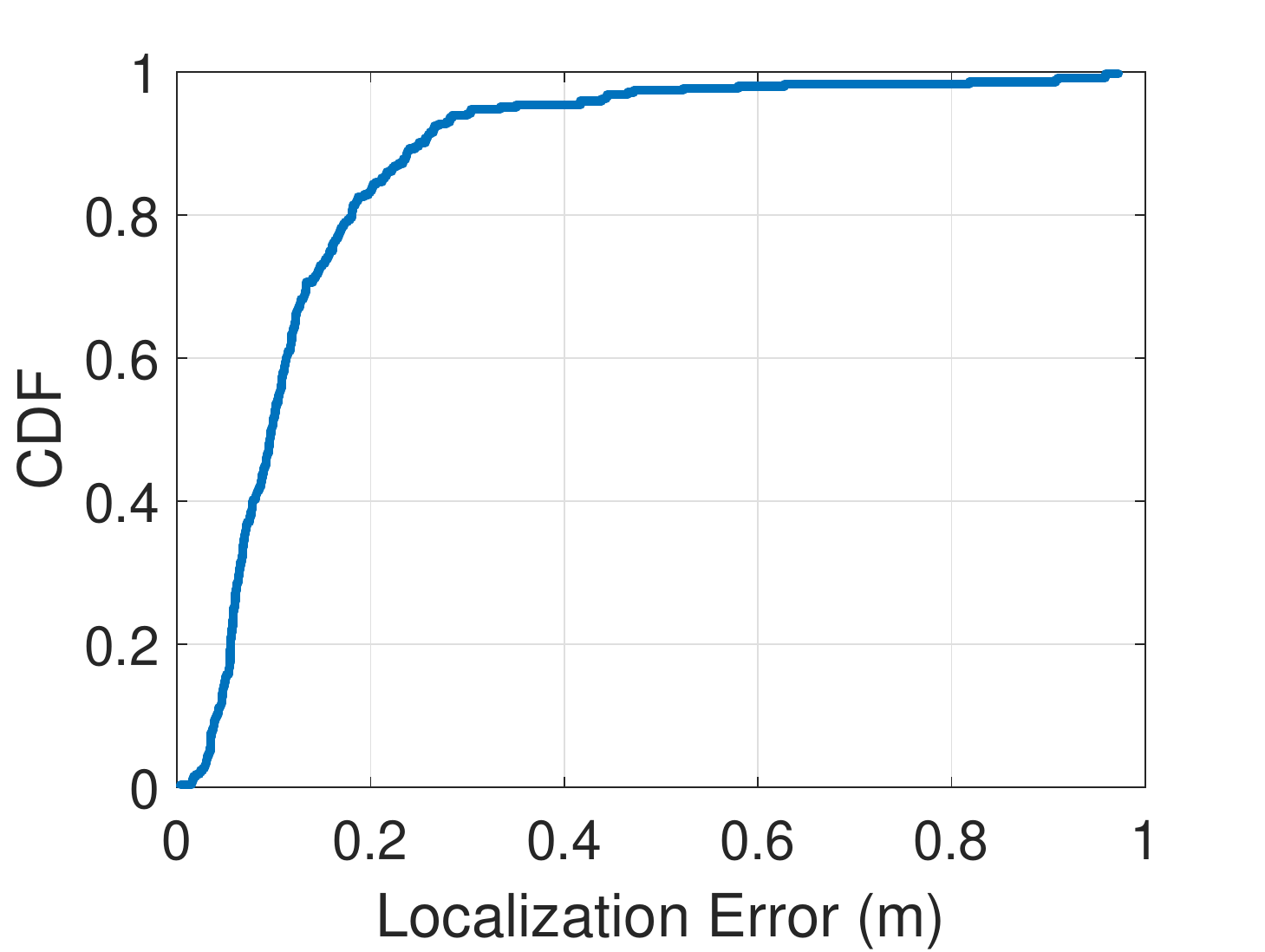}
\end{minipage}%
}%
\centering
\caption{CDF of Angle Error and Localization Error.}
\label{fig: ErrorCDF}
\end{minipage}
\end{figure*}
\begin{figure*}[t]
\begin{minipage}[t]{0.49\linewidth}
\centering
\subfigure[Angle Error]{
\begin{minipage}[t]{0.5\linewidth}
\centering
\includegraphics[width=1.6 in]{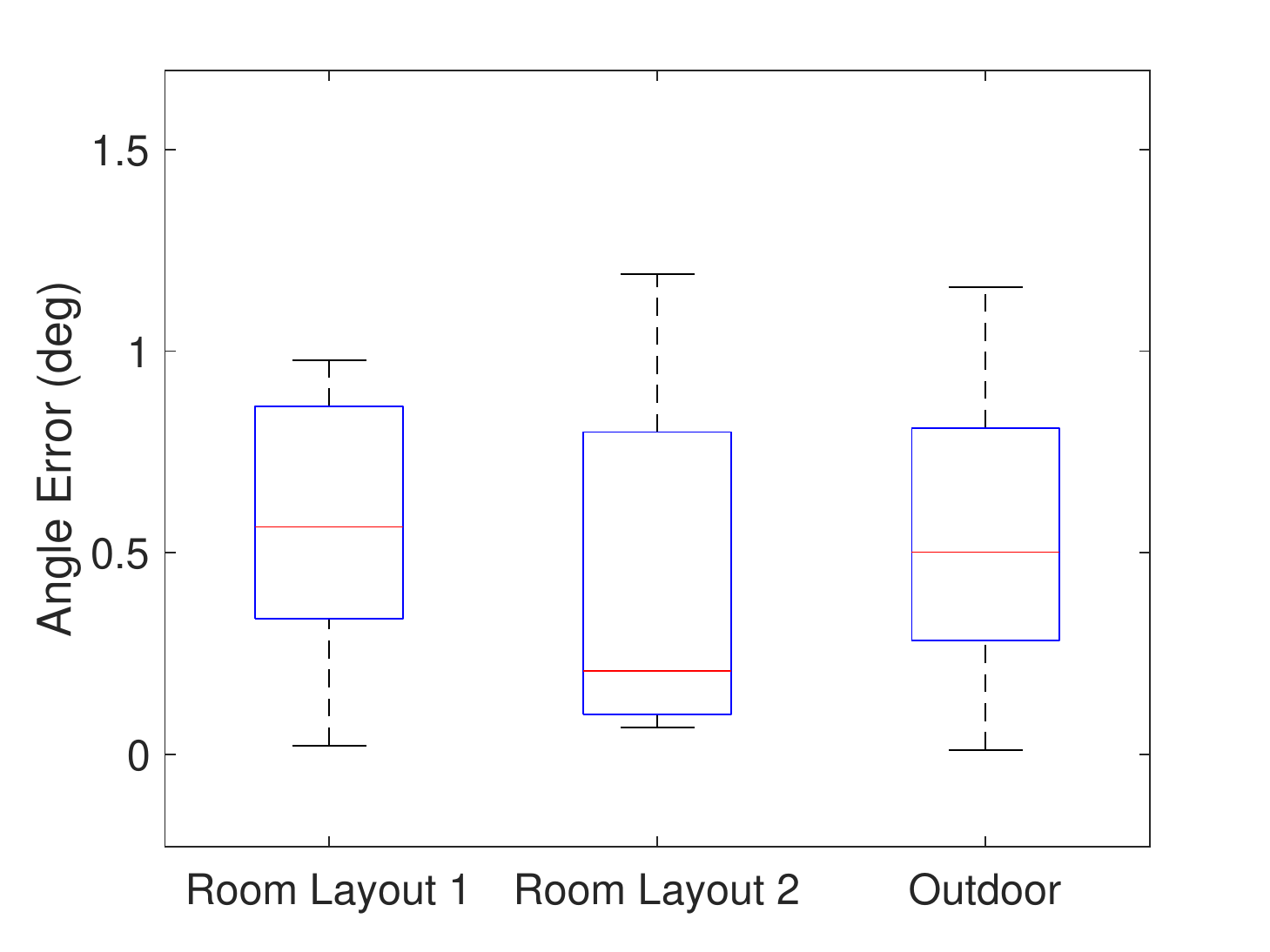}
\end{minipage}%
}%
\subfigure[Localization Error]{
\begin{minipage}[t]{0.5\linewidth}
\centering
\includegraphics[width=1.6 in]{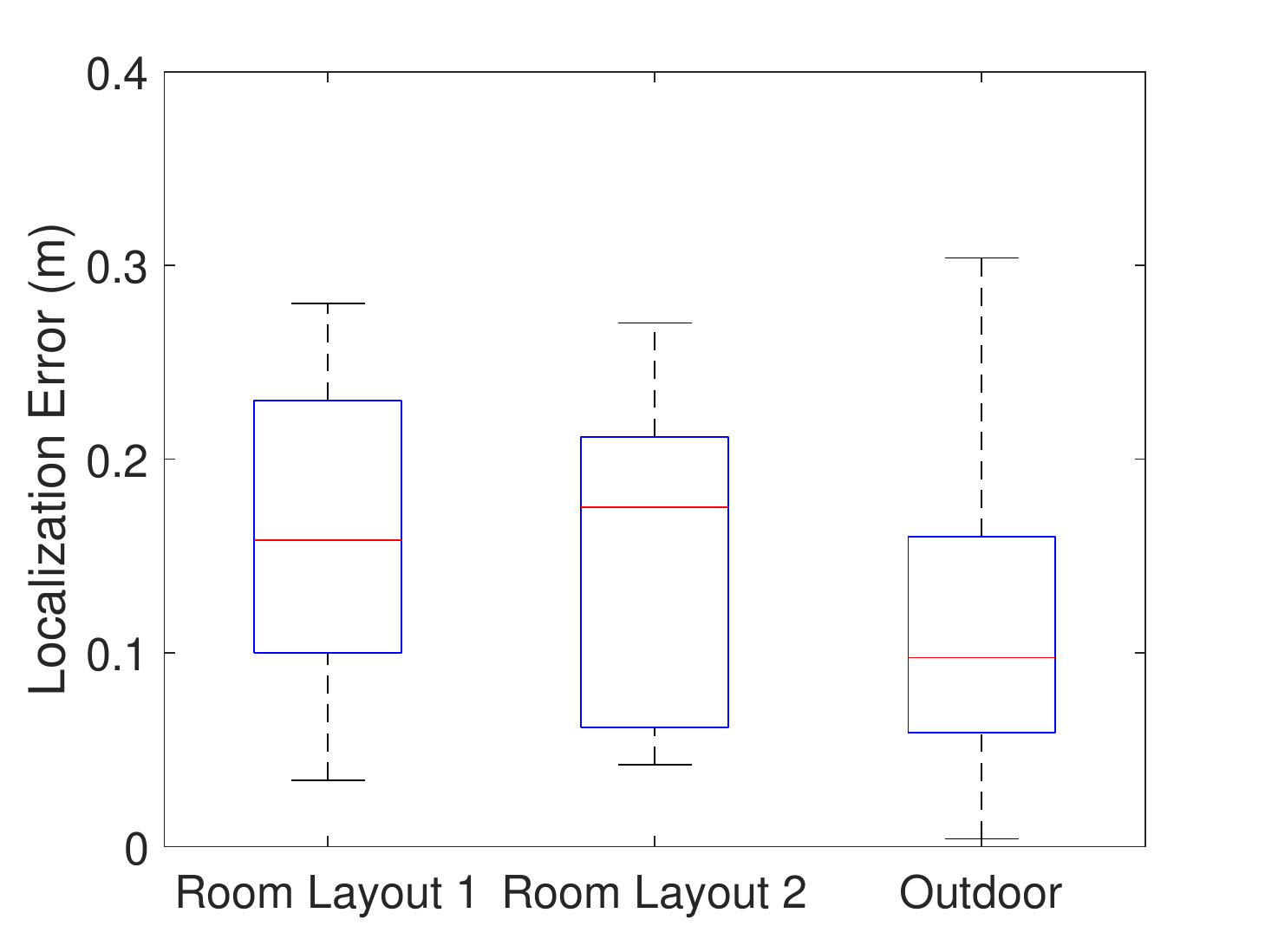}
\end{minipage}%
}%
\centering
\caption{Angle/Localization Error under different room layout.}
\label{fig: EnvImpact}
\end{minipage}
\begin{minipage}[t]{0.49\linewidth}
\centering
\subfigure[Angle Error]{
\begin{minipage}[t]{0.5\linewidth}
\centering
\includegraphics[width=1.6 in]{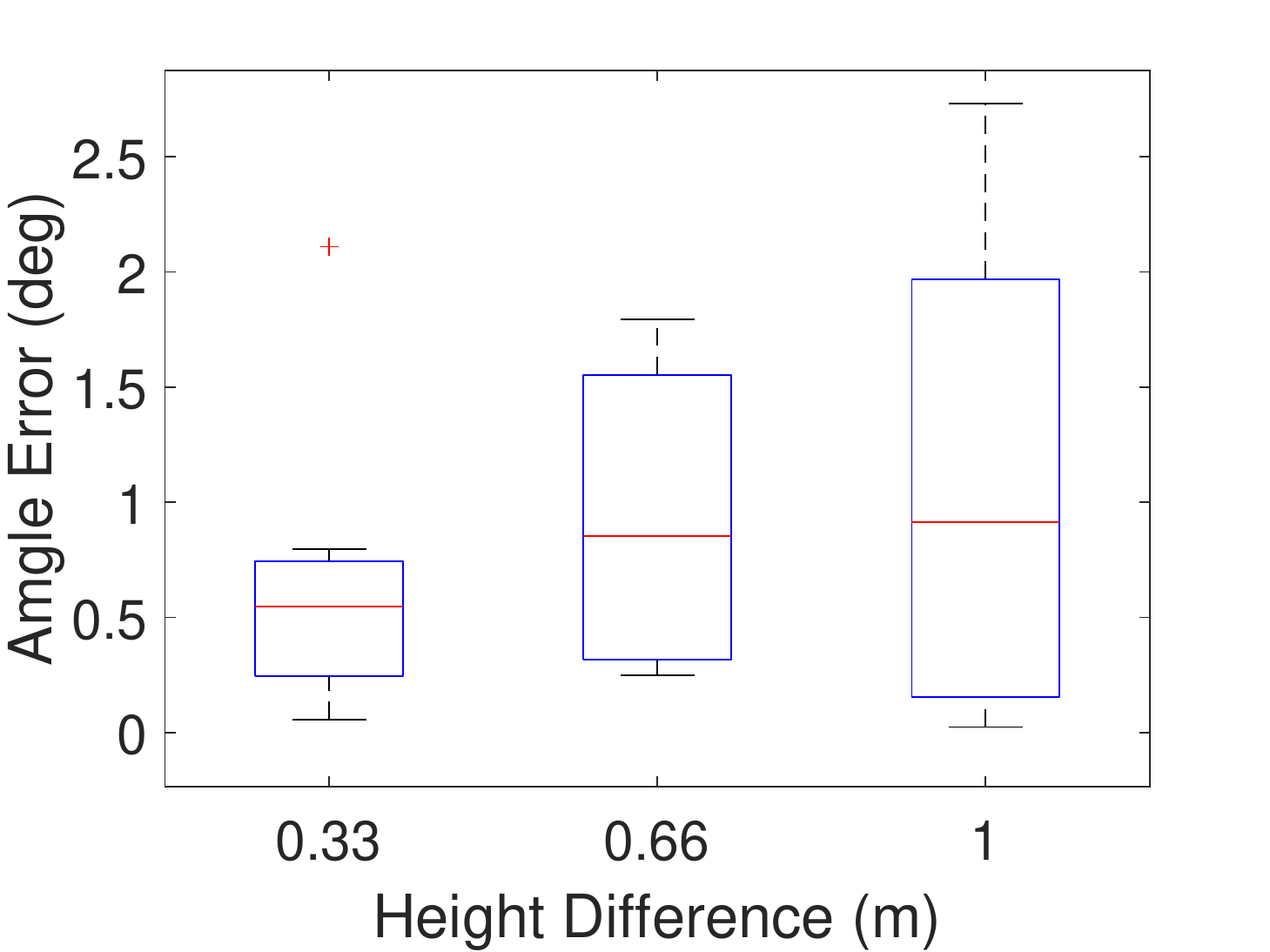}
\end{minipage}%
}%
\subfigure[Localization Error]{
\begin{minipage}[t]{0.5\linewidth}
\centering
\includegraphics[width=1.6 in]{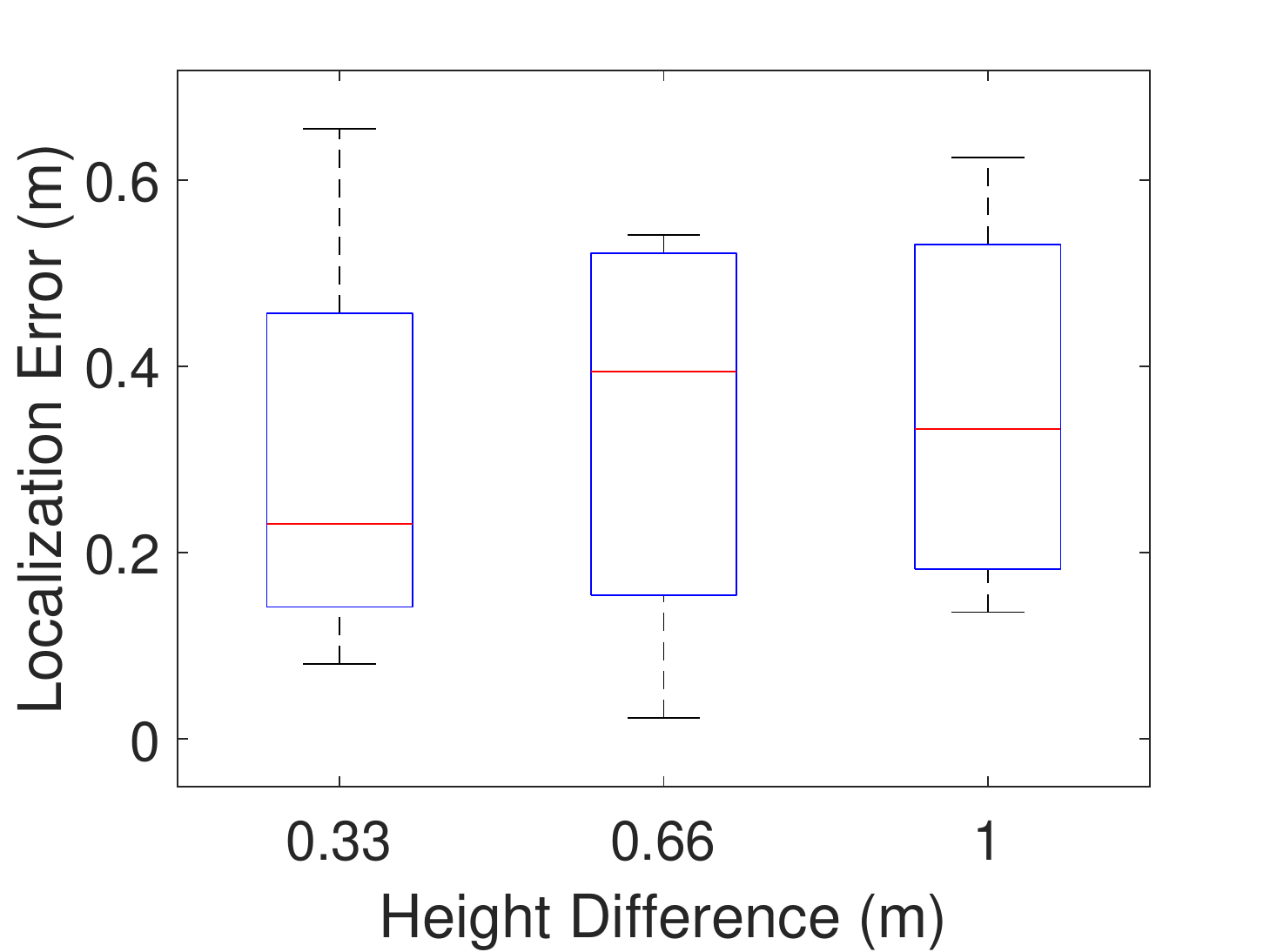}
\end{minipage}%
}%
\centering
\caption{Angle/Localization Error at different height differences between the spy radar and the detector.}
\label{fig: HeightImp}
\end{minipage}
\end{figure*}

\subsection{Performance of Spy Radar Detection}

First, we define the terminology and metrics to evaluate the performance of spy radar detection. A \textit{Spy Radar Detection} occurs when a spy radar is detected and identified as a radar. We define \textit{Spy Radar Detection Rate} to evaluate the accuracy that $Radar^2$ can correctly detect and identify a spy radar. A \textit{False Alarm} occurs when there is no spy radar while our system detects one, which comes from noise, or a WiGig device is recognized as a radar. We denote $TP, TN, FP$, and $FN$ as the numbers of True Positive, True Negative, False Positive, and False Negative. Therefore, \textit{Spy Radar Detection Rate} = $TP/(TP+FN)$ and \textit{False Alarm Rate} = $FP/(TN+FP)$.

\subsubsection{Impact of Environment}

We evaluate the spy radar detection under two different room layouts, which are shown in Fig. \ref{fig: roomlayout} and an outdoor environment. We collect 7 minutes of data for each room and outdoors. As shown in Table~\ref{tab:DetRoom}, the detection rate maintains 100$\%$ in these three environments, while the \textit{False Alarm Rate} is 4.07$\%$, 3.33$\%$ and 2.59$\%$ respectively for Room Layout 1, Room Layout 2 and outdoor environments. The result shows that the spy radar detection of $Radar^2$ will not be influenced by various environments.

\begin{table}[t]
    \centering
    \caption{Performance under various environments. (DR: Spy Radar Detection Rate, FAR: False Alarm Rate)}
    \begin{tabular}{c|c|c|c|c|c|c}
        \hline
       Environment & TP & TN & FP & FN & DR & FAR\\
       \hline
       Room Layout 1 & 270 & 261 & 9 & 0 & 100$\%$ & 4.07 $\%$\\
             Room Layout 2 & 270 & 259 & 11 & 0 & 100$\%$ & 3.33 $\%$\\
             Outdoor & 270 & 263 & 7 & 0 & 100$\%$ & 2.59 $\%$\\
        \hline
    \end{tabular}
    \label{tab:DetRoom}
\end{table}

\subsubsection{Impact of Distance and Angle}

We placed our detector at various distances and angles to evaluate their impact on the spy radar detection rate. Notice that in this experiment, we adjusted the distance and angle within the Field-of-View (FoV) of the detecting radar. As shown in Fig. \ref{fig: Detdisang} (a), the detection rate is 100$\%$ within 20m, and it will drop to 84.72$\%$ at 25m, so we conclude that the maximum range for our spy radar detection can reach 20m, which is enough for indoor environment. The FoV of the detector (TI AWR1843) is 120$\degree$, according to Fig. \ref{fig: Detdisang} (b), within the FoV of the detector, the spy radar detection rate maintains 100$\%$ under various directions. The \textit{False Alarm Rate} achieves $2.78\%$, which comes from noise in the environment.

\subsubsection{Impact of Testbeds $\&$ Types of Radar}

We evaluate the performance of $Radar^2$ under different testbeds, using different types of radar as spy radar. The spy radar detection rate achieves 100$\%$ for all cases, which reveals that $Radar^2$ is applicable to different testbeds and is able to detect different kinds of radars.

\subsubsection{Impact of Human Activity}

We evaluate our spy radar detection rate when humans walk around the room. The spy radar detection rate still reaches 100$\%$ with human activity, which shows that $Radar^2$ works well when there are human activities in the environment.

\subsubsection{Error Analysis}

The detection rate and false alarm rate are related to signal-to-noise ratio (SNR).

A free space propagation model for mmWave systems describes the relationship between distance and power received \cite{kulemin2003millimeter}:
\begin{equation}
    P_r=\frac{P_t A^2_e \sigma}{4\pi \lambda^2 d^4},
\end{equation}
where $P_r$ is the received power, $P_t$ is the transmitted power, $A_e$ is the effective surface area of antennas, $\sigma$ is a radio scattering factor and $d$ is the effective distance. We can derive the distance $d$ as:
\begin{equation}
    d=\sqrt[4]{(\frac{P_t A^2_e \sigma}{4\pi \lambda^2 P_r})}.
    \label{maxd}
\end{equation}

Equ. \eqref{maxd} indicates that within a certain range $d$, if the received power is sufficiently large $P_r>P_{th}$ or relatively large $P_r>R_{th}\cdot mean(P_r+N)$, the spy radar is able to be detected. Therefore, we can successfully detect spy radar within $d$. Based on our experiment, $d$ ranges from 20m to 25m. It is enough to be used indoors. Besides, we can increase $d$ by adjusting $R_{th}$ and $P_{th}$ for outdoor environment.

When noise is large, $Radar^2$ may mistake noise as mmWave signal, which causes a false alarm. The noise power $N$ should be comparable to our defined threshold:
\begin{equation}
    N\geq P_{th},
\end{equation}
or the noise changes dramatically:
\begin{equation}
    \frac{\max{N}}{mean(N)}>R_{th}.
\end{equation}

The probability of these two cases is small. According to our experiment, the false alarm rate is under $5\%$.

\subsection{Performance of Localization}

To evaluate the performance of spy radar localization, we first define our evaluation metrics: \textit{Angle Error} and \textit{Localization Error}. The \textit{Angle Error} is defined as the difference between the angle measured by our system and the angle measured by a goniometer. The \textit{Localization Error} is defined as the Euclidean distance between the position measured by $Radar^2$ and the position measured by a tape measure. The overall result is shown in Fig. \ref{fig: ErrorCDF}. As shown in Fig. \ref{fig: ErrorCDF} (a), 90$\%$ of \textit{Angle Error} are within 1.036 degrees, while Fig. \ref{fig: ErrorCDF} (b) shows that 90$\%$ of \textit{Localization Error} are within 0.2563m.

\subsubsection{Impact of Environment}

We evaluate that our localization system is robust in different environments, the result is depicted in Fig. \ref{fig: EnvImpact}, the \textit{Localization Error} is within 0.3 meters and the \textit{Angle Error} is within 1.5 degrees under different environments. The result indicates that the localization system in $Radar^2$ is robust to various environments.

\subsubsection{Impact of Height Difference}

$Radar^2$ provides a 2D position for spy radar localization, while the height difference between the spy radar and detector will cause errors in localization. The height of a room is usually within 3 meters, so we consider that the difference between a spy radar and our detector is within 1 meter. In this section, we will give a theoretical analysis of the impact of height and then conduct experiments to show that the height difference has a limited impact on localization accuracy.

\textbf{Theoretical Analysis:} Suppose the spy radar is located at $p_T=(x_T,y_T,z_T)$, and our detector is located at $p_i=(x_i,y_i,z_i)$. If they are placed at the same height ($z_T=z_i$), the bearing is
\begin{equation}
    \theta_i=arctan(\frac{y_i-y_T}{x_i-x_T}).
\end{equation}

When they are not placed at the same height ($z_T\neq z_i$), the bearing is
\begin{equation}
    \theta_i'=arctan(\frac{y_i-y_T}{\sqrt{(x_i-x_T)^2+(z_i-z_T)^2}}).
\end{equation}

The difference between $\theta_i$ and $\theta_i'$ is negligible if the height difference $z_i-z_T$ is small enough. To verify it, we simulate the AoA error within 10 meters in the 2D plane when the height difference is 1 meter.

\begin{figure}[t]
    \centering
    \includegraphics[width=150pt]{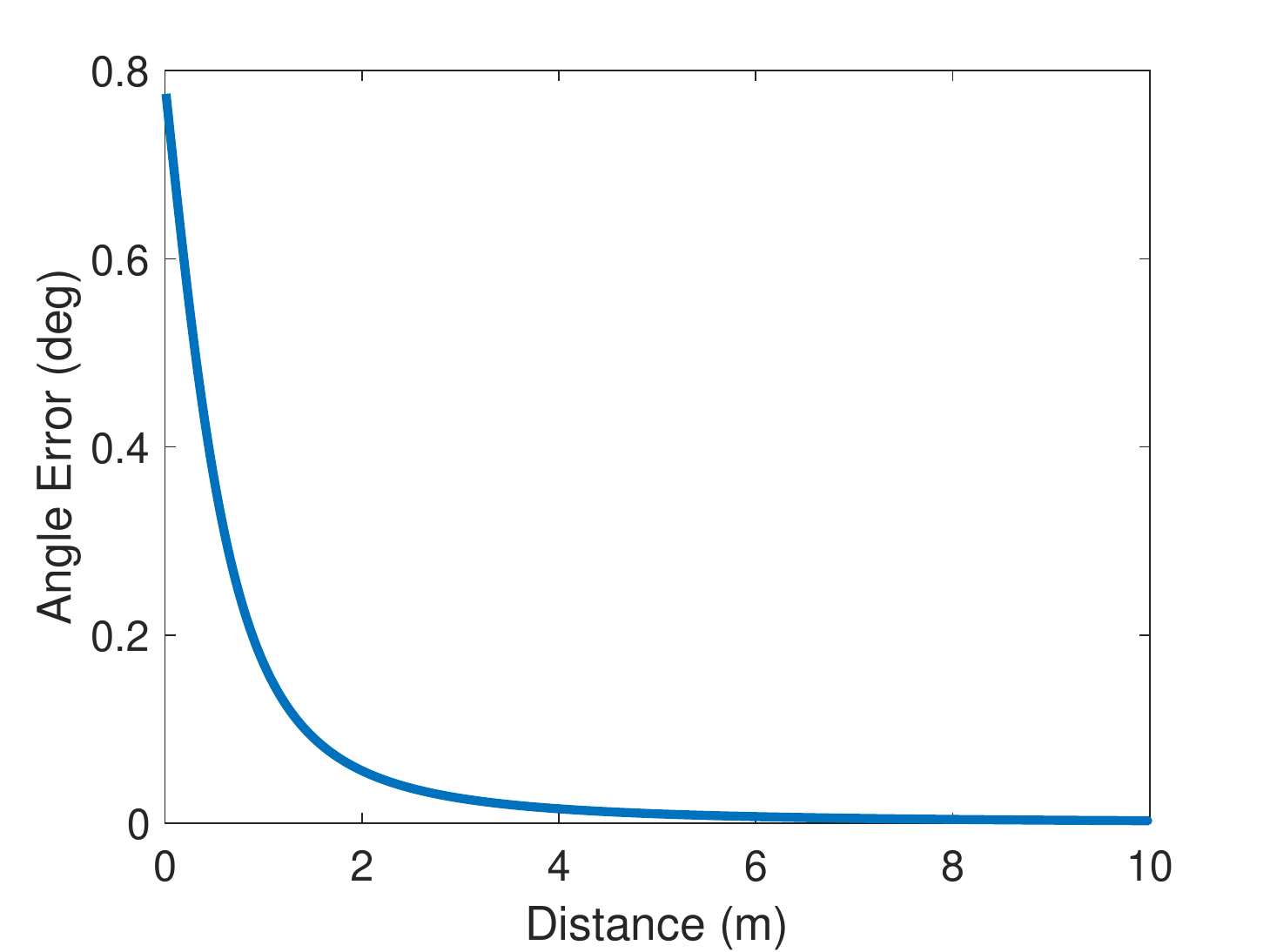}
    \caption{Theoretical Error in AoA estimation when the height difference of the spy radar and the detector is 1m.}
    \label{fig:ThereoErr}
\end{figure}

\textbf{Simulation Result:} As shown in Fig. \ref{fig:ThereoErr}, with increasing distance in the 2D plane, the \textit{Angle Error} introduced by the height will be smaller. Within a 1-meter height difference, the \textit{Angle Error} is within 0.8 degrees, which is acceptable for \textit{Triangulation}.

\textbf{Experiment Result:} By adjusting the height difference to 0.33m, 0.66m, and 1m, the \textit{Angle Error} and \textit{Localization Error} is shown in Fig. \ref{fig: HeightImp}. The result reveals that the \textit{Angle Error} is within 2.5 degrees and the \textit{Localization Error} is within 0.7m when the height difference is within 1m. Thus we don't need to consider the height difference between devices at home.

\subsubsection{Impact of Number of Anchors}

As we propose the \textit{nearest approach} for spy radar localization, the localization accuracy depends on the number of anchors. We evaluate the impact of the number of anchors by moving our detector to more anchor points. AoA is estimated at each anchor independently, so \textit{Angle Error} is not related to the number of anchors. We will compare the \textit{Localization Error} with different numbers of anchors.

\begin{figure}[t]
    \centering
    \includegraphics[width=150pt]{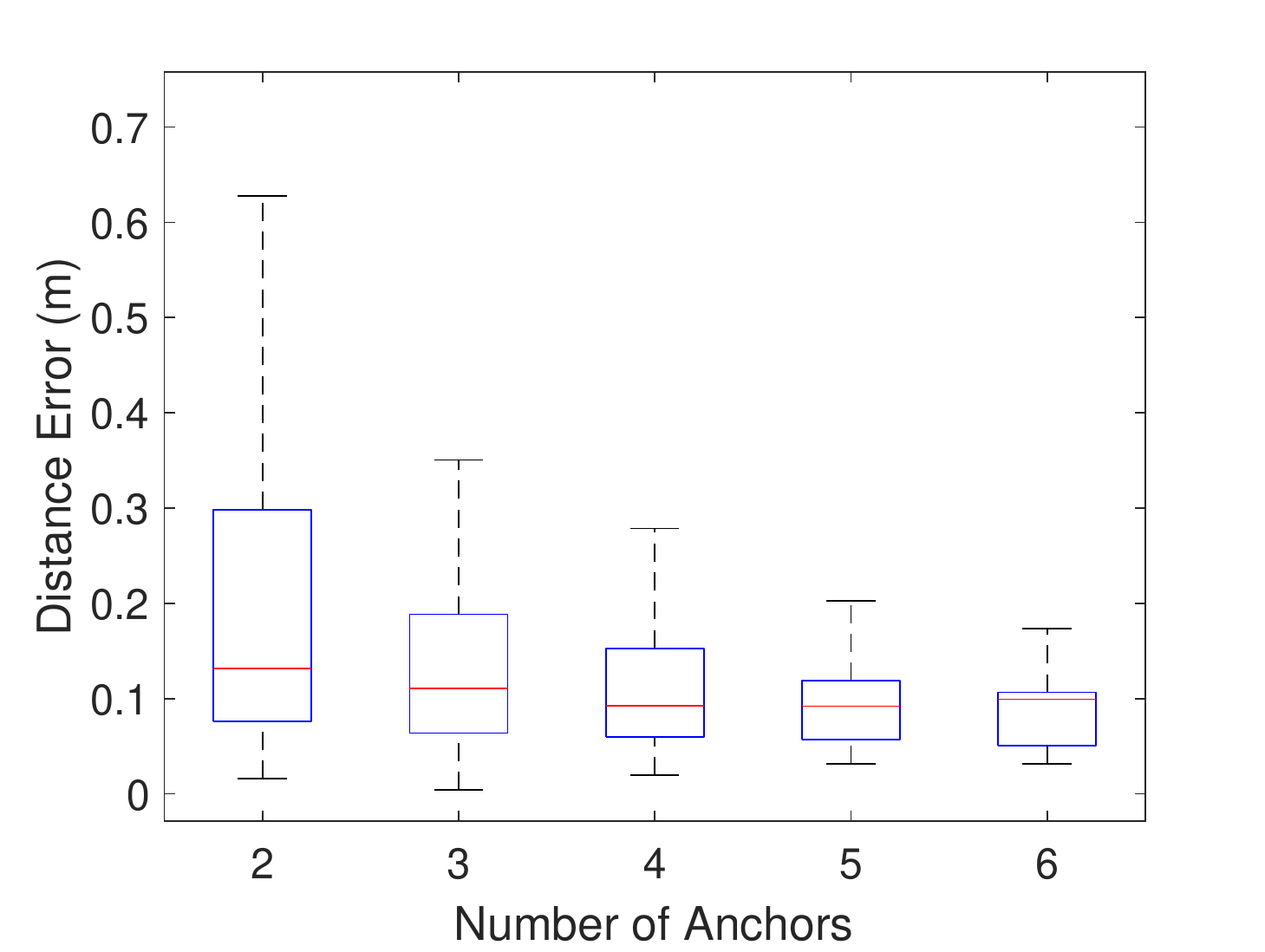}
    \caption{Localization Error vs. the number of anchors.}
    \label{fig:NoAnchor}
\end{figure}

As shown in Fig. \ref{fig:NoAnchor}, the \textit{Localization Error} decreases with an increasing number of anchors. However, when the number of anchors reaches five, the \textit{Localization Error} will not decrease significantly. This experiment instructs us to choose a suitable number of anchors, which should balance the localization performance with the cost of the system.

\subsection{Performance of Multi-Device Scenario}

To evaluate $Radar^2$ when multiple mmWave devices work simultaneously, we place two radars working in different modes and use a WiGig device as an interference term. 

\subsubsection{Spy Radar Detection}

The overall \textit{Spy Radar Detection Rate} achieves 96.67$\%$ in this experiment. Table~\ref{tab:DeteTwoRadar} depicts the spy radar detection result when two radars work simultaneously, while when a spy radar and a WiGig device coexist, the spy radar detection rate maintains $100\%$. $Radar^2$ sometimes cannot detect one of the spy radars when two radars work simultaneously. The reason is that these two radars work in different modes (e.g., CW mode and FMCW mode). The energy of FMCW sometimes dominates that of CW. Besides, if the frequency of CW is within the frequency range of FMCW, then the detector cannot detect CW radar. However, this situation can be improved by removing one detected radar and re-detecting spy radars since localization can help us find the location of the spy radar.

\begin{table}[t]
    \centering
    \caption{Spy Radar Detection Rate when two radars coexist.}
    \begin{tabular}{c|c}
    \hline
 Result & Detection Rate\\
 \hline
 Two Radars are Detected & 94.44$\%$\\
       One Radar is Detected & 5.56$\%$\\
       No Radar is Detected & 0$\%$\\
       \hline
    \end{tabular}
    \label{tab:DeteTwoRadar}
\end{table}

\subsubsection{Localization}

\begin{figure}[t]
    \begin{minipage}[htb]{1\linewidth}
        \centering
        \subfigure[Angle Error]{
            \begin{minipage}[t]{0.48\linewidth}
            \centering
            \includegraphics[width=1.6 in]{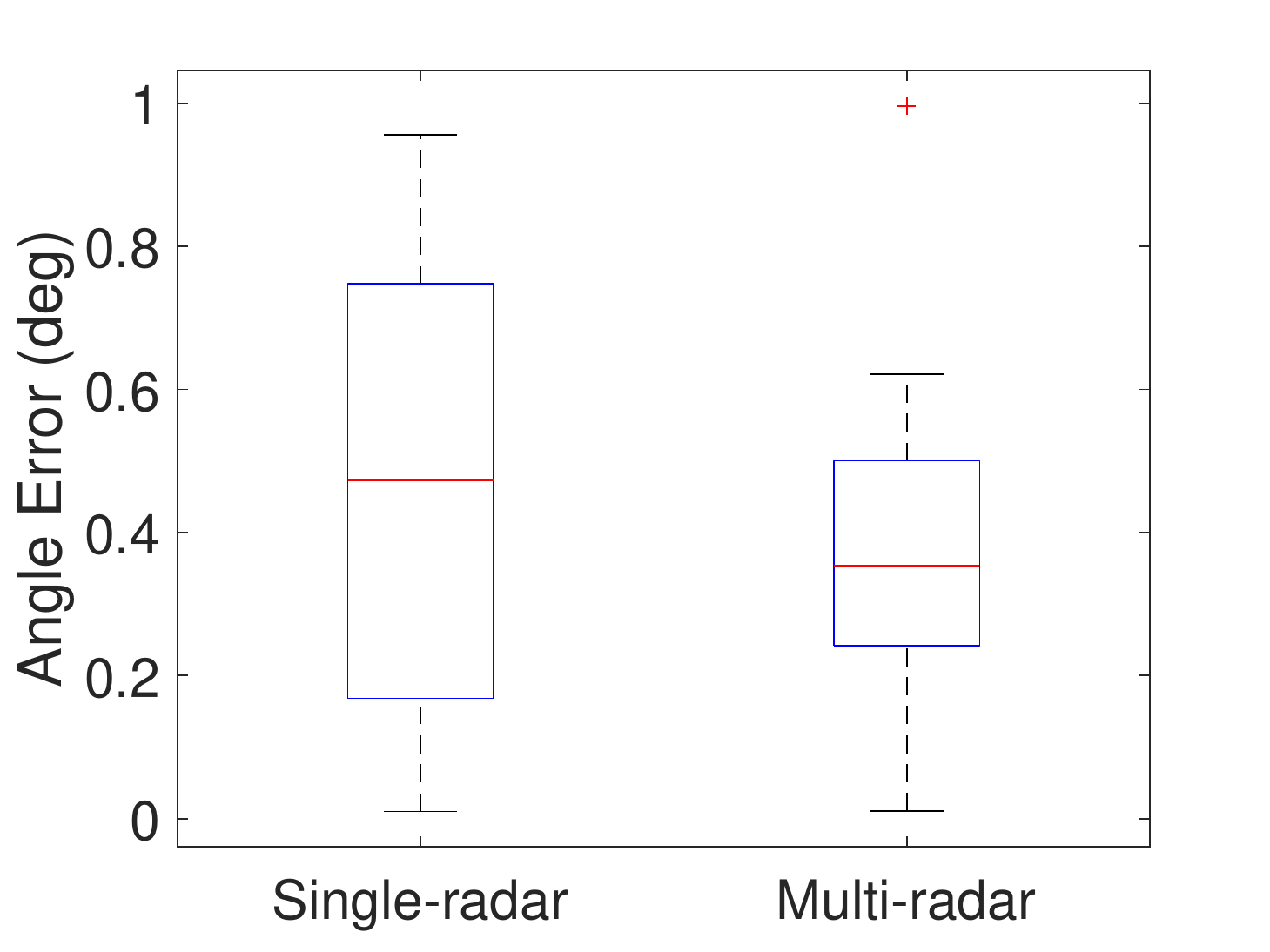}
            \end{minipage}%
        }%
        \subfigure[Localization Error]{
            \begin{minipage}[t]{0.48\linewidth}
            \centering
            \includegraphics[width=1.6 in]{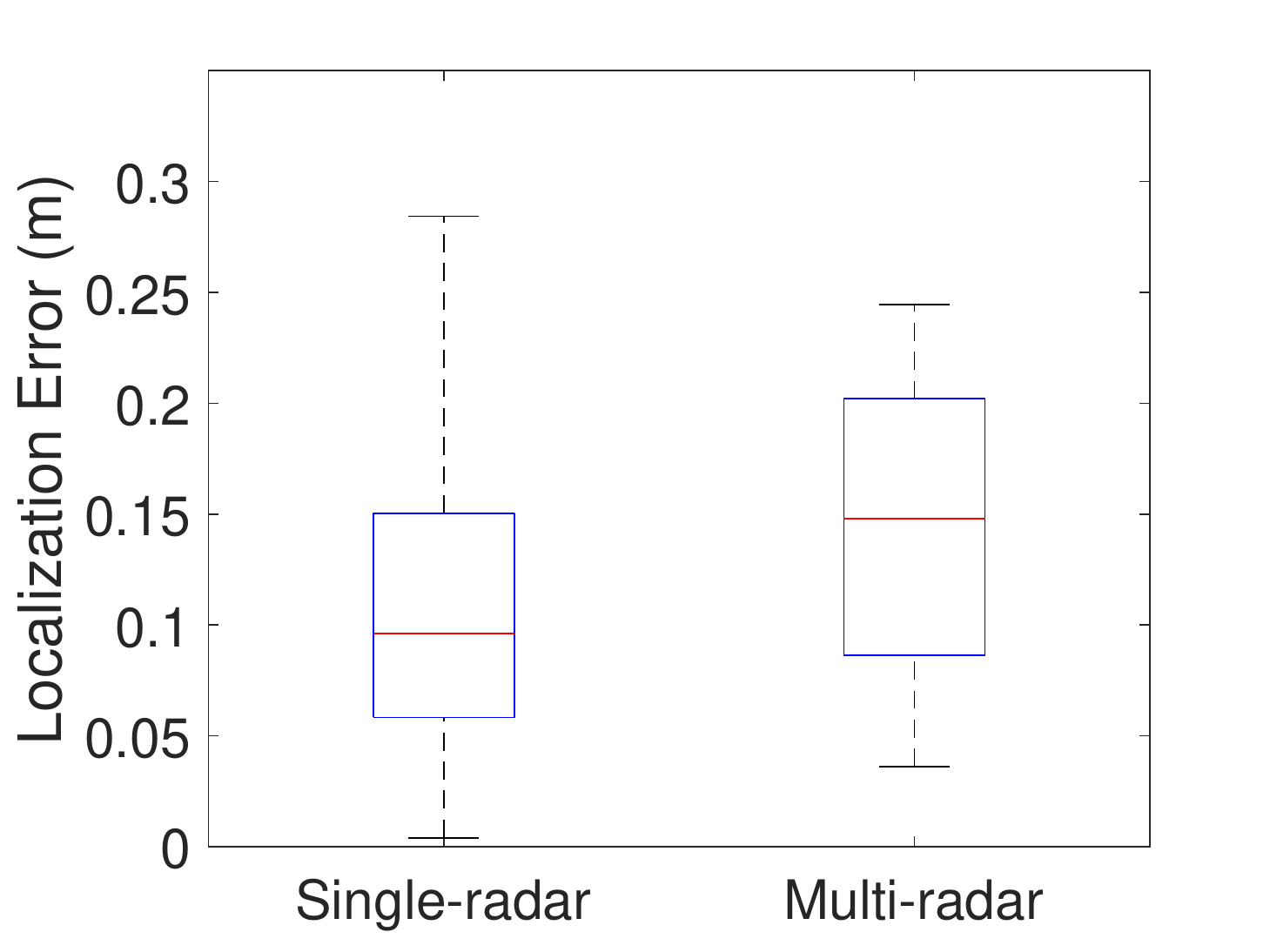}
            \end{minipage}%
        }%
        \centering
        \caption{Angle/Localization Error in the multi-device scenario.}
        \label{fig: MulLoc}
    \end{minipage}
\end{figure}

In Fig. \ref{fig: MulLoc}, the \textit{Angle Error} and the \textit{Localization Error} perform well under multi-device scenario: the \textit{Angle Error} is within 0.7 degrees while the \textit{Localization Error} is within 0.3m. The result reveals that $Radar^2$ can adapt well to the scenario where multiple mmWave devices coexist.

\section{Discussion}

In this section, we discuss the application scenarios and limitations of $Radar^2$.

\textbf{Extension to other Frequency Band.}
In our system, we specialize in detecting and localizing spy radars in the mmWave band. Meanwhile, many other radars are working on other frequency bands ~\cite{yue2020bodycompass,fan2020home}. Moreover, our architecture can be easily applied to detect and localize radars operating in other frequency bands. A spy radar can be detected by using a radar that works in the same band, then generating single-frequency signals and a sweep-frequency signal to demodulate from the received signal. As the principle and processing method are almost the same, we did not implement the system in other frequency bands. Besides, as a mmWave radar usually has a smaller size compared with other low-frequency band radars, it is easier to hide it as a spy radar.

\textbf{Extension to 3D Localization.}
$Radar^2$ is designed to detect and localize spy radars in 2D spaces. While 2D positioning can solve most cases at home, it may not be sufficient at an airport or large shopping mall, where height plays an important role in positioning. Our design can also be extended to enable 3D positioning. In that case, the anchors should be deployed in 3D space, which means that the detector should be moved to different heights for more observations. With more measurements and applying \textit{Triangulation} on the 3D dimension, 3D localization can also be achieved.

\textbf{WiGig Sensing.}
In $Radar^2$, we assume that WiGig is used for communication. However, WiGig can also be used to sense the environment ~\cite{hong2020wbf}. Hence, WiGig has the potential to spy on people's daily lives. In our $Radar^2$ system, we did not categorize WiGig as a spy radar. However, $Radar^2$ can detect the existence of a WiGig transceiver. Therefore, to detect a spy WiGig, users can turn off their own WiGig devices and then use our system to detect the presence of WiGig. If a WiGig transceiver still exists, it must be a spy WiGig, which can be further localized and removed.

\section{Related Work}

In this section, we summarize related works on device detection, waveform classification, and device localization.

\textbf{Device Detection.} Prior works proposed to detect the RF device by unintended electromagnetic emissions (UEEs) with a spectrum analyzer \cite{acharya2013detection, weng2005neural, acharya2012system}. Shikhar et al. proposed that UEEs can be taken as a unique signature of electronic devices. This signature can be used for device detection and identification. In addition, they designed a model that can accurately detect if there is a device nearby emitting UEE or not \cite{acharya2012system}. They also showed a novel approach to the application of Principal Component Analysis (PCA) in detecting UEEs \cite{acharya2013detection}. Weng et al. designed a neural network for automated device detection using UEEs \cite{weng2005neural}. However, to obtain UEEs, an expensive spectrum analyzer is required, which is unaffordable for daily use.

Besides, nonlinear effects were explored to recognize hidden electronics \cite{mazzaro2014detection, li2018eye}. Li et al. proposed E-Eye, which leveraged mmWave to recognize hidden electronics using nonlinear effect \cite{li2018eye}. However, to support nonlinear effects in mmWave, they require a specialized radar, and they are not able to localize such devices.

\textbf{Waveform Classification.} Most existing works were motivated to classify different communication signals for channel sensing, and spectrum allocation \cite{fehske2005new, shi2019deep, soltani2019real, caffrey1995space}. Shi et al. studied deep learning-based signal classification for wireless networks in the presence of out-network users and jammers \cite{shi2019deep}. Soltani et al. designed a real-time and embedded RF signal classifier \cite{soltani2019real}. Their work only involved communication signals, and mmWave band signals were not considered.

Different machine learning methods were explored to classify civil and military radars \cite{petrov2013radar, chen2017radar, revillon2018radar}. Petrov et al. designed a neural network for timely and reliable recognition of radar signal emitters \cite{petrov2013radar}. Chen et al. proposed a new framework to classify radar emitters for large data set \cite{chen2017radar}. However, no work has differentiated communication signals (WiGig) from radar signals in 60GHz. In other words, previous work didn't differentiate communication-based RF signals and sensing-based RF signals.

\textbf{Device Localization.} Even though mmWave radars can localize objects (e.g., human tracking) \cite{zeng2016human,sengupta2020mm,zhao2021human}, they cannot identify which object it is. Therefore, localizing a mmWave radar is a new problem that was not studied by others. 

Multiple prior works proposed localizing a smartphone by fingerprinting \cite{azizyan2009surroundsense, farshad2013microscopic} or RFID tags \cite{wahl2013using}. However, these technologies cannot be used to localize spy radars as we cannot train the spy radar's information through fingerprinting. Nguyen et al.\cite{nguyen2019towards} proposed \textit{Triangulation} to locate drones at different anchors, which brings insight for us to locate a spy radar.

\section{Conclusion}
In this paper, we propose $Radar^2$, a passive mmWave Detection and Localization system using COTS mmWave radar. As an innovation, we first design a \textit{Frequency Component Detection} method to detect the existence of mmWave signal by comparing frequency components in the received signal and carrier signals. Besides, to differentiate WiGig from radar signals, we leverage the spectrum as features for waveform classification. Finally, we provide a spy radar localization method that utilizes triangulation. We implemented $Radar^2$ in different kinds of mmWave radar testbeds. Experimental results show that $Radar^2$ is able to detect various types of radar working in various frequency bands. Our system, $Radar^2$, reaches up to 96\% spy radar detection rate and 0.3m localization error within 20 meters without COTS mmWave hardware modification, and it performs well under various scenarios.

\section*{Acknowledge}

The authors would like thank the anonymous reviewers and
the Associate Editor for providing constructive and generous feedback.

\appendix
\section{Appendix}
\label{SLo}
We provide mathematical analysis for spy radar localization in Appendix.

Suppose we have $N$ anchor points $p_1=(x_1,y_1)$, $p_2=(x_2,y_2)$, ..., $p_N=(x_N,y_N)$, at each anchor we estimate the AoA of the spy radar: $\theta_1,\theta_2$, ..., $\theta_N$. The objective is to estimate the position of the spy radar $p_T=(x_T, y_T)$. We have the following equations:
\begin{equation}
\begin{split}
    \frac{x-x_1}{\sin(\theta_1)}&=\frac{y-y_1}{\cos(\theta_1)}=a_1,\\
    \frac{x-x_2}{\sin(\theta_2)}&=\frac{y-y_2}{\cos(\theta_2)}=a_2,\\
    &\dots\\
    \frac{x-x_N}{\sin(\theta_N)}&=\frac{y-y_N}{\cos(\theta_N)}=a_N,\\
\label{linear}
\end{split}
\end{equation}
where each equation is a line represented by an anchor position with the AoA of the spy radar estimated at this anchor. Thus, the $N$-line linear system can be expanded in the following way:
\begin{equation}
\begin{split}
     \left\{\begin{array}{lr}
    x+0\times y - \sin(\theta_1)\times a_1 - 0 \times a_2 \dots -0 \times a_N = x_1\\
    0\times x+ y - \cos(\theta_1)\times a_1 - 0 \times a_2 \dots -0 \times a_N = y_1\\
    x+0\times y - 0\times a_1 - \sin(\theta_2) \times a_2 \dots -0 \times a_N = x_2\\
    0\times x+ y - 0\times a_1 - \cos(\theta_2) \times a_2 \dots -0 \times a_N = y_2\\
    \dots\\
    x+0\times y - 0\times a_1 - 0 \times a_2 \dots -\sin(\theta_N) \times a_N = x_N\\
    0\times x+ y - 0\times a_1 - 0 \times a_2 \dots -\cos(\theta_N) \times a_N = y_N\\
    \end{array}
    \right.
\label{mline}
\end{split}
\end{equation}

Accordingly, the matrix representation in Equ. \eqref{mline} can be written as:
\begin{equation}
    \textbf{Gm}=\textbf{d},
\end{equation}
where $\textbf{G}$ is a matrix of size $(N \times 2) \times (N + 2)$:
\begin{equation}
\label{G}
    \textbf{G}=\begin{bmatrix}
    1 & 0 & -\sin(\theta_1) & 0 &       & 0 \\
    0 & 1 & -\cos(\theta_1) & 0 &       & 0 \\
    1 & 0 & 0 & -\sin(\theta_2) & \dots & 0 \\
    0 & 1 & 0 & -\cos(\theta_2) &       & 0 \\
      &  &  \vdots &  & \ddots&\vdots \\
    1 & 0 & 0 & 0 & \dots & -\sin(\theta_N) \\
    0 & 1 & 0 & 0 &       & -\cos(\theta_N)
     \end{bmatrix},
\end{equation}
and $\textbf{m}$ and $\textbf{d}$ are two column vectors,
\begin{equation}
    \textbf{m}=\begin{bmatrix}
    x &
    y &
    a_1 &
    a_2 &
    \dots &
    a_N
     \end{bmatrix}^T,
\end{equation}
\begin{equation}
    \textbf{d}=\begin{bmatrix}
    x_1 &
    y_1 &
    x_2 &
    y_2 &
    \dots &
    x_N &
    y_N
     \end{bmatrix}^T.
\end{equation}

The objective is to find $\textbf{m}$ which minimize $||\textbf{Gm-d}||^2$,
\begin{equation}
    \hat{\textbf{m}}=\arg_{\textbf{m}}\min(||\textbf{Gm-d}||^2).
\end{equation}
The position of the spy radar should be the first two elements in $\hat{\textbf{m}}$:
\begin{equation}
    \hat{p}_T=(\hat{x}_T,\hat{y}_T)=\hat{\textbf{m}}(1:2).
\end{equation}

To solve such a least square (LS) problem, we use SVD decomposition:
\begin{equation}
    \textbf{G}=\textbf{U}_{(N\times 2)\times(N+2)}\times \textbf{S}_{(N\times 2)\times(N+2)} \times \textbf{V}^T_{(N\times 2)\times(N+2)}.
\end{equation}

We expand the above SVD representation of $\textbf{G}$ in terms of the columns of $\textbf{U}$ and $\textbf{V}$, and simplify it into the compact forms:
\begin{equation}
\begin{split}
    \textbf{G}&=\begin{bmatrix}
        \textbf{U}_k & \textbf{U}_0
    \end{bmatrix}\begin{bmatrix}
        \textbf{S}_p & \textbf{0} \\
        \textbf{0} & \textbf{0}
    \end{bmatrix}\begin{bmatrix}
        \textbf{V}_k & \textbf{V}_0
    \end{bmatrix}^T, \\
    \textbf{G}&=\textbf{U}_k\times \textbf{S}_k \times \textbf{V}^T_k.
\end{split}
\end{equation}

Obtaining the SVD solution, the optimal $\textbf{m}$ is:
\begin{equation}
    \hat{\textbf{m}}=\textbf{V}_k\times \textbf{S}^{-1}_k\times \textbf{U}^T_k \times \textbf{d}.
\end{equation}

The SVD solution gives the point nearest to all $N$ lines given in 2D space:
\begin{equation}
    \hat{p}_T=(\hat{x}_T,\hat{y}_T)=\hat{\textbf{m}}[1:2].
\end{equation}

Besides, we get the error $\epsilon$:
\begin{equation}
\begin{split}
    \epsilon= \sum_i [ (\hat{x}_T-\hat{\textbf{m}}(2+i) \cdot \sin(\theta_i)-x_i)^2+ \\ (\hat{y}_T-\hat{\textbf{m}}(2+i) \cdot \cos(\theta_i)-y_i)^2], 
\end{split}
\end{equation}
which defines the Euclidean lengths from the nearest point to all lines.

\bibliographystyle{IEEEtran}
\bibliography{reference}

\vfill

\end{document}